\documentclass[reprint,pra,superscriptaddress,longbibliography,amsmath,amssymb,aps] {revtex4-1}
\usepackage{graphicx}
\usepackage{dcolumn}
\usepackage{bm}
\usepackage{afterpage}
\usepackage{float}
\usepackage{placeins}
\usepackage{dsfont}
\usepackage[usenames,dvipsnames,svgnames,table]{xcolor}
\usepackage[utf8]{inputenc}
\usepackage{bm}
\usepackage{amssymb}
\usepackage{amsmath}
\usepackage{amsfonts}
\usepackage{graphicx}
\usepackage{dcolumn}
\usepackage{bm}
\usepackage[pdftex,
            pdftitle={LGTSU2_U1},
            pdfauthor={Pablo Sala},
            bookmarks,
            colorlinks,
            linkcolor=black,
            citecolor=magenta,
            menucolor=black,
            urlcolor=blue,
            plainpages=false,
            pdfpagelabels,
            hypertexnames=false]{hyperref}

 \renewcommand{\v}[1]{\ensuremath{\mathbf{#1}}}

\newcommand{\abs}[1]{\left| #1 \right|}
\newcommand{\avg}[1]{\left< #1 \right>}

\newcommand{\ket}[1]{\left| #1 \right>}

\let\baraccent=\= \renewcommand{\=}[1]{\stackrel{#1}{=}}

\begin{document}
\title{Variational study of U(1) and SU(2) lattice gauge theories with Gaussian states in 1+1 dimensions}

\author{P. Sala}
\thanks{where $^*$ means co-first authors.}
\affiliation{Department of Physics, T42, Technische Universit{\"a}t M{\"u}nchen, James-Franck-Stra{\ss}e 1, D-85748 Garching, Germany}
\email{pablo.sala@tum.de}

\author{T.Shi}
\thanks{where $^*$ means co-first authors.}
\affiliation{Institute of Theoretical Physics, Chinese Academy of Sciences, P.O. Box 2735, Beijing 100190, China}
\email{tshi@itp.ac.cn}

\author{S. K{\"u}hn}
\affiliation{Perimeter Institute for Theoretical Physics, 31 Caroline Street North, Waterloo, ON N2L 2Y5, Canada}

\author{M. C. Ba{\~n}uls}
\affiliation{Max-Planck-Institut f\"ur Quantenoptik, Hans-Kopfermann-Straße 1, 85748 Garching, Germany}
\author{E. Demler}
\affiliation{Department of Physics, Harvard University, Cambridge, Massachusetts 02138, USA}

\author{J. I. Cirac}
\affiliation{Max-Planck-Institut f\"ur Quantenoptik, Hans-Kopfermann-Straße 1, 85748 Garching, Germany}
\date{\today }

\begin{abstract}

We introduce a method to investigate the static and dynamic properties of both Abelian and non-Abelian lattice gauge models in 1+1 dimensions. Specifically, we identify a set of transformations that disentangle different degrees of freedom, and apply a simple Gaussian variational ansatz to the resulting Hamiltonian. To demonstrate the suitability of the method, we analyze both static and dynamic aspects of string breaking for the U(1) and SU(2) gauge models. We benchmark our results against tensor network simulations and observe excellent agreement, although the number of variational parameters in the Gaussian ansatz is much smaller.
\end{abstract}

\maketitle




%




\section{Introduction}

Gauge theories lie at the basis of our fundamental understanding of nature. Quantum electrodynamics (QED) describes the interactions of electrons and positrons with the electromagnetic field and is based on the U(1) gauge symmetry group. Quantum chromodynamics (QCD) accounts for the strong interaction between quarks and gluons and is based on the SU(3) gauge group. Those theories are roots of the Standard Model, that comprises our current understanding of particle physics. While in perturbative limits they are very well understood \cite{Weinberg}, this is not the case in general.

A very powerful framework to address nonperturbative regimes is lattice gauge theory (LGT) \cite{Wilson,Kogut79} where space (and time) is discretized. In such a theory the fermionic (matter) degrees of freedom reside in the sites of a cubic lattice, the bosonic (gauge) ones in the links, and they interact with each other in a gauge invariant way. The continuum limit is then recovered when the lattice constant is taken to zero, by properly renormalizing the coupling constants in the process. Monte Carlo methods~\cite{Aoki2017,QCDPH,Monte} have successfully been used in LGT to compute with a very high precision several physical properties in different models. This approach works extremely well as long as the so-called sign problem \cite{signpro} is absent, which is the case for static (thermal equilibrium) problems in QED or QCD in the absence of a chemical potential.

In dynamical scenarios or regimes in which Monte Carlo simulations suffer from the sign problem, one has to look for other techniques. Hence, there is an ongoing interest in overcoming these limitations~\cite{Cristoforetti2013,Hebenstreit2013a,Scorzato2015,Ammon2016,Gattringer2016}.

Hamiltonian lattice methods, as pioneered in Refs.~\cite{Banks1976,Hamer1977,Banks1977,Crewther1980,Hamer1982a,Horn1984,Horn1985,Morningstar1992,Hollenberg1994,Morningstar1996a}, might offer another possibility. Recently, several research groups have addressed relatively simple lattice gauge models in the Hamiltonian formulation using tensor network techniques~\cite{Orus2014,Verstraete2008}, motivated by the success of DMRG~\cite{White1992} and related approaches to solve strongly correlated condensed matter systems in lattices.

Those methods are based on variational ans{\"a}tze over families of states. Most of the work~\cite{Hamer2,Banuls,CiracStudy,Buyens2016,VQED,Banuls2015,Banuls2016,kuhnEE,Buyens2014,Buyens2017a,Milsted2016,Zapp2017,Rico2014} so far has been concentrated in 1+1 dimensions, where such family corresponds to Matrix Product States (MPS). Despite their simplicity, those models contemplate many of the phenomena that are expected to occur in higher dimensions, such as confinement~\cite{SusSch}, string breaking~\cite{VQED}, etc. In fact, some studies have successfully analyzed models with the sign problem \cite{U1Zoller,kuhn,Brenes2018,Buyens2017,CiracStudy}, and thus raise the expectations about tensor network methods complementing Monte Carlo techniques. However, the extension of these methods to higher dimensions is still under development, and it is not clear if the methods will succeed in such cases (other methods suggesting to use projected entangled pair states for the study of lattice gauge theories have been proposed \cite{Zohar2015,Zohar2016,Zohar2018,Zapp2017}). Another important class of states that are commonly used for variational calculations are Gaussian states \cite{Bravyi2005,Peschel2009,Kraus2009,Bravyi20052,Weedbrook2012}. Those are defined for bosonic and fermionic theories, and comprise all states that can be generated by a Gaussian function of creation and annihilation operators acting on the vacuum. As they fulfill Wick's theorem, one can compute expectation values very efficiently and thus use them for variational calculations \cite{Lieb1981,Bach1994,Kraus,Tao}. Unfortunately, in the case where both bosons and fermions are present, Gaussian states cannot accommodate any correlations between them beyond mean-field, and thus they are not useful for the description of lattice models with matter and gauge fields. Apart from that, their special form makes them  unsuitable to study many complex phenomena.

In this paper we show how one can use Gaussian states as variational ansatz for LGT in 1+1 dimensions to study both ground state and dynamical properties. The main idea is to first apply transformations that disentangle the bosonic and fermionic degrees of freedom, followed by transformations that convert the Gaussian states in suitable ans{\"a}tze for variational calculations. A similar procedure has been recently successfully applied to condensed matter models \cite{Tao,TaoKondo}. We apply such method to the U(1) (Schwinger model) and SU(2) gauge groups, with special emphasis on the latter. We analyze the ground state, as well as the time dynamics in the presence of external charges. In order to benchmark the approach, we compare our results to those from either published MPS calculations~\cite{VQED, U1Zoller} or by explicitly performing MPS simulations ourselves. Despite the fact that the number of variational parameters in the Gaussian ansatz is much lower than in MPS, we observe very good agreement, thereby showing the suitability of our approach to LGT simulations. Furthermore, the method presented here does not suffer from a violation of the area law, in contrast to MPS, which makes it suitable also for the study of real time dynamics after quenches.

Using the transformed Hamiltonian, we study static as well as dynamic properties of  string breaking. First, to probe the static aspects of the phenomenon, we compute the interacting vacuum of the theory in the presence of external charges. This allows us to determine the static potential, i.e. the excess energy compared to the vacuum without external charges as a function of the distance between the external charges \cite{Polyakov}. We demonstrate that the method reliably distinguishes between the regimes where a flux string is present in the ground state and string breaking occurs. Second, we simulate the real-time evolution of a flux string, compute local observables and monitor the spatially resolved flux profiles as well as correlation functions throughout the evolution. These studies allow us to clearly distinguish between the string and broken string cases in both the ground state and dynamical evolution. The ansatz captures the relevant features and it is possible to simulate the dynamics even in the scenario of a global quench.

The rest of the paper is structured as follows. In Sec. \ref{sec:model} we briefly review the Kogut-Susskind Hamiltonian lattice formulation \cite{Susskind} for a gauge theory with a compact symmetry group.
Afterwards we show how the gauge field can be decoupled for systems with open boundary conditions (OBC) and discuss how to apply the variational method to the resulting formulation in Sec. \ref{sec:methods}.
Once the general framework has been established, we benchmark our approach for two specific cases. In Sec. \ref{sec:U1} we test the ansatz for describing static properties (Sec. \ref{sec:U1static}) and real-time dynamics of string breaking (Sec. \ref{sec:realU1}) for a U(1) LGT. For the former, we compare our results to those obtained in Ref.~\cite{VQED} close to the continuum limit. In Sec. \ref{sec:SU2} we turn to the non-Abelian case of a SU(2) LGT. We introduce two additional unitary transformations which allow us to apply the Gaussian variational ansatz in the presence of external charges by decoupling dynamic and static fermions (Sec. \ref{sec:DecVV}). Again, we characterize the ground state (Sec. \ref{Sp}) and real-time dynamics (Sec. \ref{sec:reSU2}) of the system and benchmark our results against MPS simulations. Finally, we conclude in Sec. \ref{sec:out}.

\section{Model\label{sec:model}}
The model we are studying is a (1+1)-dimensional LGT with compact gauge group. We adopt the Kogut-Susskind Hamiltonian formulation with staggered fermions~\cite{Susskind} which reads
\begin{align}
\begin{aligned}
H& =\varepsilon\sum_{n=1}^N\left( {\bm{\phi}_{n}^{\dagger }U_{n}^{j} \bm{\phi}_{n+1}+\mathrm{H.c.}}\right)    \\
& +m \sum_{n=1}^N{(-1)^{n}\bm{\phi}_{n}^{\dagger }\bm{\phi}_{n}}+\frac{g^2}{2}\sum_{n=1}^{N-1}{\bm{L}_{n}^{2}}
\end{aligned}
\label{HU}
\end{align}
on a lattice with $N$ sites. For Eq.~\eqref{HU} to correspond to the discretization of the continuum theory as in Ref.~\cite{Susskind}, one identifies $\varepsilon=1/(2a)$, $g^2=g^2_0a$ and $m=m_0$, where $g_0$ and $m_0$ are the bare coupling and mass, and $a$ is the lattice spacing. In general, for compact gauge groups $G$ the operators $U_n^j$ are unitaries $U_{n}^{j}=\exp \big({i\sum_a^{\text{dim}(G)}{\theta _{n}^{a }T^{a,j}}}\big)$
with as many independent angular variables $\theta ^{a}$ as generators of the Lie algebra associated to $G$, $T^{a ,j}$. These generators satisfy
\begin{equation}\label{CR}
\lbrack T^{a ,j},T^{b ,j}]=i\sum_{c }{f^{a b c}T^{c,j}}
\end{equation}%
where $f^{abc }$ are the structure constants, and $j$ labels a given representation. These angular variables $\theta^a _{n}$ are related to the gauge field on a link $n$ as $\theta^c_n=-agA^{1,c}_n$. The fermionic field $\phi_n$ is a spinor in the same representation as $U^j$ and resides on site $n$. The electric term in Eq.~\eqref{HU} can be written in terms of either the left or right electric fields $\bm{L}_{n}$ and $\bm{R}_{n}$ where each of them has $\text{dim}(G)$ components. They are related by a group element in the adjoint representation $U_{n}^{\text{Adj.}}$, $\bm{R}_{n}=U_{n}^{\text{Adj.}} \bm{L}_{n}$, and fulfill the commutation relations
\begin{align}
[R_{a },R_{b }]=i\sum_{c }{f^{abc }R_{c}},\quad
[L_{a },L_{b }]=-i\sum_{c}{f^{abc}L_{c }}.
\end{align}
Moreover, as conjugate momenta of the gauge field, $\bm{L}_{n}$
and $\bm{R}_{n}$ fulfill the commutation relations%
\begin{align}
\begin{aligned}
& [L_{k}^{a},(U_{k}^{j})_{\alpha\beta}]=\sum_{\gamma}{(T^{a,j})_{\alpha\gamma}(U_{k}^{j})_{\gamma\beta}}  \\
& [R_{k}^{a },(U_{k}^{j})_{\alpha\beta}]=\sum_{\gamma}{(U_{k}^{j})_{\alpha \gamma}(T^{a,j})_{\gamma \beta}}.
\end{aligned}
\label{eq:commutation_LU_RU}
\end{align}

Physical states $|\psi\rangle$ have to fulfill Gauss' law, $G^a_n|\psi\rangle = 0$ for all $a,n$, where
\begin{equation}
G_{n}^{a }:=L_{n}^{a }-R_{n-1}^{a }-\mathcal{Q}_{n}^{a}.
\label{GU}
\end{equation}
In the expression above $\bm{\mathcal{Q}}_{n}$ is the total charge which consists of the dynamical $\bm{Q}_{n}$ and static (external) charges $\bm{q}_{n}$ at the site $n$, $\bm{\mathcal{Q}}_{n}=\bm{Q}_{n}+\bm{q}_{n}$.

In the case of the Abelian group U(1), $\bm{\phi}_{n}$ is just a single component fermionic field and, since the structure constants vanish, $\bm{R}_{n}=\bm{L}_{n}=L_{n}$. The link operators reduce to $U_n=\exp(i\theta_n)$, where the phase $\theta_n\in[0,2\pi]$ represent an Abelian phase related to the gauge field as $\theta_n=-agA_n^1$. In this case the commutation relations between the conjugate variables from Eq.~\eqref{eq:commutation_LU_RU} yield
\begin{equation}
[\theta_n,L_m]=i\delta_{n,m}.
\end{equation}
The staggered charge is defined as $Q_{n}=\phi _{n}^{\dagger }\phi_{n}-(1-(-1)^{n})/2$, and $q_{n}$ is simply a real number. In the limit of strong coupling, $g\gg 1$, meaning that the hopping term in Eq.~\eqref{HU} can be neglected, the Hamiltonian can be solved analytically. The gauge invariant ground state in the sector of vanishing total charge is simply given by the odd sites occupied by a single fermion, empty even sites and the links carrying no electric flux
\begin{align}
|\psi_{\text{SC},\text{U(1)}}\rangle = |\mathbf{1};\mathbf{0};\mathbf{1};\mathbf{0};\dots \rangle \otimes |0\rangle_\text{gauge}.
\label{eq:SC_U1}
\end{align}
In the expression above the numbers in bold face indicate the fermionic occupation and $|0\rangle_\text{gauge}$ indicates the total electric flux carried by gauge links. This state is the lattice analog of the Dirac sea or the bare vacuum of the theory.

For the gauge group SU(2), the fermionic fields in the fundamental $j=1/2$ representation are given by %
\begin{equation}
\bm{\phi}_{n}=(\phi _{n}^{r},\phi _{n}^{g})^{T},\hspace{15pt}\bm{\phi}%
_{n}^{\dagger }=(\phi _{n}^{r,{\dagger }},\phi _{n}^{g,{\dagger }})
\end{equation}%
taking into account the two colors components $\phi ^{r}$ (``red'') and $\phi ^{g}$ (``green''). The structure constants are given by the completely antisymmetric Levi-Civita symbol $f^{abc}=\epsilon^{abc}$ and the generators are represented by $T^{a,1/2}=\sigma^a/2$ with $\sigma^a$ the Pauli matrices. Therefore there are three independent angular variables $\theta^{a}$ on each link.

The total SU(2) color charge is then given by $\bm{\mathcal{Q}}_{n}=\bm{Q}_n+\bm{q}_n$ with three different components
\begin{equation}
{Q}_{n}^{a}=\frac{1}{2}\bm{\phi}_{n}^{\dagger }\sigma ^{a }\bm{\phi}%
_{n},\hspace{15pt}q_{n}^{a}=\frac{1}{2}\sigma ^{a }
\end{equation}%
for $a=x,y,z$, where the external charges at a given site $\bm{q}_n$ are nothing but spin operators. Note that when there is no static charge on site $n$ then $q^a_n=0$.

Similar to the Abelian case of U(1), the ground state in the limit of large $g$ can be solved analytically yielding the bare vacuum
\begin{align}
 |\psi_{\text{SC},\text{SU(2)}}\rangle = |\mathbf{1,1};\,\mathbf{0,0};\,\mathbf{1,1}\dots\rangle \otimes |0\rangle_\text{gauge}
 \label{eq:SC_SU2}
\end{align}
where the bold numbers now correspond to the occupation numbers of each color of fermions on a site and $|0\rangle_\text{gauge}$ again indicates the gauge links carrying no color flux.

Unless stated otherwise, we fix for all the following the hopping amplitude to $\varepsilon=1$ as unit of energy. Moreover, we also set $\hbar=1$.

\section{Methods\label{sec:methods}}
\subsection{Decoupling the gauge field \label{sec:decgau}}
Because of the absence of transversal directions in 1+1 dimensions, the gauge degrees of freedom are not truly independent. Hence, it is always possible to decouple the matter and the gauge fields by applying a unitary transformation to the Hamiltonian \eqref{HU}. Here we present a simple way of performing such a decoupling for systems with OBC. While similar transformations were carried in \cite{NAtrans,Lenz1994} for periodic boundary conditions, the resulting Hamiltonian takes a simpler form in our case. This unitary transformation works for any gauge symmetry given by a compact Lie group and a unitary representation. In the following we briefly summarize the main steps and show the full derivation in Appendix \ref{app:Elim}.

The decoupling is achieved with the unitary transformation%
\begin{equation}
\,\Theta =\prod_{k=1}^{\rightarrow }{\exp \Big(i \bm{\theta}_{k}\cdot
\sum_{m>k}{\bm{\mathcal{Q}}_{m}\Big)}},
\label{thetamod}
\end{equation}%
where the superscript $\rightarrow $ means that the product must be ordered from left to right with increasing site index $k$.
Applying this transformation to the Hamiltonian, we obtain
\begin{align}
\begin{aligned}
H_{\Theta }:=\Theta H\Theta ^{\dagger}& =\varepsilon \sum_{n}\left( {\bm{\phi}_{n}^{\dagger } \bm{\phi}_{n+1}+}\mathrm{H.c.}\right)  \\
& +m \sum_{n}{(-1)^{n}\bm{\phi}_{n}^{\dagger }\bm{\phi}_{n}}+\frac{g^2}{2}H_{e}
\end{aligned}
\label{H}
\end{align}%
where the electric energy term in the sector of vanishing total charge exhibits a long-range Coulomb interaction
\begin{equation}
H_{e}=\sum_{a }{\sum_{n,m}{\mathcal{Q}_{n}^{a }V_{n,m}\mathcal{Q}%
_{m}^{a }}}  \label{He}
\end{equation}%
between the charges with $V_{n,m}=-\left\vert n-m\right\vert /2$. Moreover, in the sector of vanishing total charge Gauss' law is transformed to $L_{n}^{a}=\Theta ^{\dagger }G_{n}^{a}\Theta =0$ acting on the physical space.

A few comments are in order. The transformation shown above is completely general and does not rely on the Gaussian variational approach. For the case of U(1) the resulting Hamiltonian is equivalent to the one used in previous numerical studies~\cite{Hamer,Banuls,CiracStudy} and recently realized in a quantum simulation experiment~\cite{Exp}. Hence, the transformed Hamiltonian from Eq.~\eqref{H} might be suitable for both the design of future quantum simulators as well as for other numerical methods.

\subsection{Variational approach \label{sec:varmet}}

In order to solve the transformed Hamiltonian $H_\Theta$, we apply a time-dependent variational method following the ideas from Ref.~\cite{Tao}. Our variational ansatz in the untransformed frame corresponding to the Hamiltonian \eqref{HU} is given by
\begin{equation}
\ket{\psi}=\Theta^{\dagger}U_{\text{ext}}\ket{\text{GS}} { \left\vert 0\right\rangle _{\text{gauge}}}
\end{equation}
where $\ket{\text{GS}}$ is a general fermionic Gaussian state, $\Theta$ is the unitary transformation from Eq.~\eqref{thetamod}, and $U_{\text{ext}}$ is another unitary transformation which decouples dynamic and static fermions (see Sec. \ref{sec:DecVV} for more details).  In particular, we see that the transformation $\Theta$ is not Gaussian. As a result, although $\ket{\text{GS}}$ is a Gaussian state, the ansatz $|\psi\rangle$ in the original frame is not.

Our goal is to study the evolution of $|\psi\rangle$ under the Hamiltonian \eqref{HU} in either imaginary or real time to obtain, respectively, the ground state or the dynamic properties. Equivalently, we can study the evolution of $\ket{\text{GS}}$ under the rotated Hamiltonian $U_{\text{ext}}\Theta H \Theta^\dagger U_{\text{ext}}^\dagger$. As shown for example in Refs.~\cite{Bach1994,Kraus,Tao}, every fermionic Gaussian state is completely characterized by its covariance matrix
\begin{equation}
\Gamma= \begin{pmatrix}
 \left\langle \bm{\phi}\bm{\phi}^{\dagger} \right\rangle &
 \left\langle \bm{\phi}\bm{\phi} \right\rangle \\
 \left\langle \bm{\phi}^{\dagger}\bm{\phi}^{\dagger} \right\rangle &
 \left\langle \bm{\phi}^{\dagger}\bm{\phi} \right\rangle \\
\end{pmatrix}
\end{equation}
which collects all two-point correlation functions  since $\bm{\phi}$, $\bm{\phi}^{\dagger}$ stand for the vectors collecting all annihilation and creation operators respectively, i.e., $\bm{\phi}=(\bm{\phi}_1,\bm{\phi}_2,\dots,\bm{\phi}_M)^T$ with $M$ the number of modes. For the Schwinger model $M=N$, whereas for the SU(2) LGT we have twice as many modes, $M=2N$, due to the two colors of fermions. Thus, in order to compute the evolution of $|\psi\rangle$, we have to determine the evolution of $\Gamma$ in imaginary and real time which is given by the equations~\cite{Tao}
\begin{align}
\frac{d}{d\tau }\Gamma (\tau )& =\{\Gamma ,\mathcal{H}(\Gamma )\}-2\Gamma
\mathcal{\ H}(\Gamma )\Gamma   \label{DynIm} \\
i\frac{d}{dt}\Gamma (t)& =[\mathcal{H}(\Gamma ),\Gamma ].  \label{DynRe}
\end{align}
In the expression above $\mathcal{H}(\Gamma )$ is the (effective) single-particle Hamiltonian in the $U_{\text{ext}}\Theta$-rotated frame (see Appendices \ref{app:Elim} and \ref{sec:SU2eff} for details). Eq.~\eqref{DynIm} yields the ground state in the limit $\tau \to \infty$, while Eq.~\eqref{DynRe} describes real-time dynamics in the family of fermionic Gaussian states.

\section{U(1) gauge theory \label{sec:U1}}
Let us first consider the simple Abelian case of the Schwinger model. Since for the U(1) gauge group the external charges are merely complex numbers, and hence $U_{\text{ext}}=\hat{\mathds{1}}$, the variational ansatz reads  $\ket{\psi}=\Theta^{\dagger}\ket{\text{GS}}\left\vert 0\right\rangle _{\text{gauge}}$. We will first analyze the (interacting) ground state (subsection A), and then the real-time dynamics in the presence of two static charges (subsection B). In both cases we will consider the decoupled Hamiltonian from Eq.~\eqref{H} that contains the matter fields only. In order to benchmark the Gaussian variational approach, we compare our results to the MPS computations carried out in Refs.~\cite{VQED,U1Zoller}.

\subsection{Static properties \label{sec:U1static}}
In this section, we study the static potential between external charges for both the massless and the massive Schwinger model. First we compute the ground state energy $E_{\text{vac}}$ by evolving the bare vacuum (Dirac sea) from  Eq.~\eqref{eq:SC_U1} according to Eq.~\eqref{DynIm}. In order to determine the ground state energy $E_Q(L)$ in the presence of static charges, we repeat the calculation but this time placing on top of the vacuum a pair of external charges separated by a distance $L$ and connected by a string of electric flux(to fulfill Gauss' law).

\begin{figure*}[tbp]
\begin{center}
\includegraphics[width=0.99\linewidth]{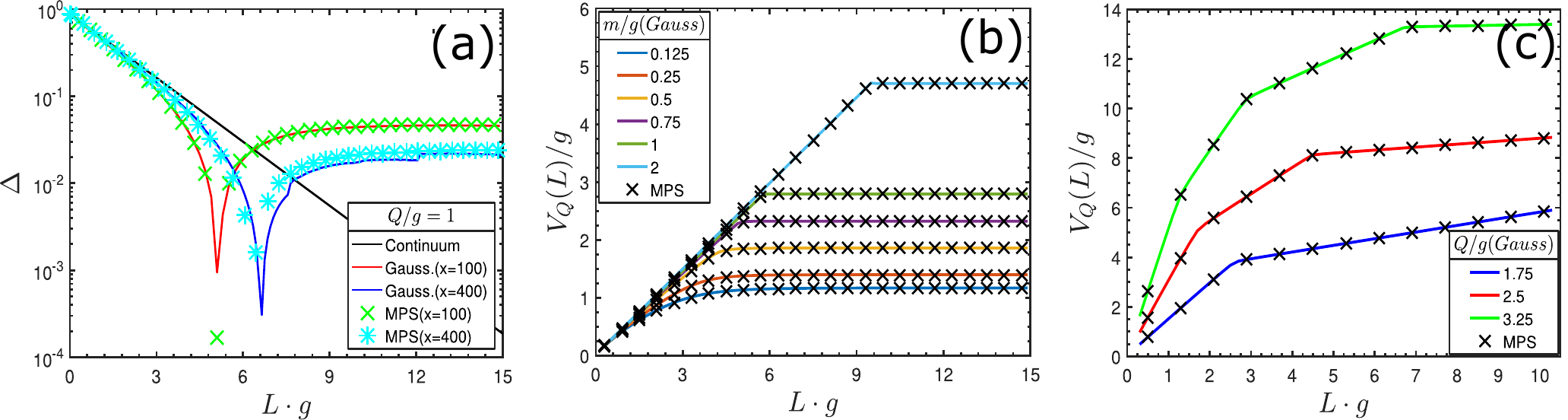}
\end{center}
\caption{Comparison between the static potential $V_Q(L)/g$ obtained from the Gaussian method and the MPS calculation from Ref.~\cite{VQED}. (a) Deviation from the continuum analytical result for the massless case $m/g=0$ (black solid line) for $x=100$ (red solid line for the Gaussian ansatz and green crosses for the MPS) and $x=400$ (blue solid line for the Gaussian ansatz and cyan asterisks for the MPS). (b) Static potential for the massive case with $Q/g=1$ and $x=100$ from the Gaussian ansatz (solid lines) and the MPS results (crosses). (c) Partial string breaking for noninteger values of $Q/g$ and $m/g=1$, $x=400$. Solid lines again represent the result from the Gaussian ansatz, crosses the MPS results from Ref.~\cite{VQED}.}
\label{fig1}
\end{figure*}

In Fig.~\ref{fig1}, we compare our results for the static potential, $V_Q(L)=E_Q(L)-E_{\text{vac}}$, to those obtained via a gauge-invariant MPS simulation in Ref.~\cite{VQED} where the gauge field was not eliminated. To compare our results to those in  Ref.~\cite{VQED}, we identify $\varepsilon=1/(2a)$ and $g^2=g^2_0a$. Then the continuum limit of the model is obtained as $a\to0$. One can introduce the dimensionless parameter $x=1/(g_0a)^2$, such that the continuum limit corresponds to $x\to\infty$.

In Fig.~\ref{fig1}(a) we show numerical results from MPS and Gaussian ans{\"a}tze in the case of two small but finite lattice spacing values, for the quantity
\begin{equation}
\Delta =\frac{1}{g}\left\vert V_{Q}(\infty )-V_{Q}(L)\right\vert
\end{equation}
where $V_{Q}(\infty )$ is the exact analytical value of $V_Q(L)$ when $L\to \infty$ in the continuum limit \cite{Iso}. We observe that both ans{\"a}tze behave very similarly, and the relative error between both is below $0.3\%$ considering MPS as the reference result.

For the massive Schwinger model, we choose $x=100$ in order to compare the potential $V_{Q}(L)$ with that in Ref.~\cite{VQED}. Although this case is not exactly solvable, it is well-known \cite{Adam,VQED} that as long as the static charges are integer multiples of the fundamental charge $g$, these will be completely screened by the particle-antiparticle pairs created out of the vacuum in the broken string case. As soon as $L$ reaches the critical distance $L_{c}$ and this happens, the potential $V_{Q}(L)$ saturates to a constant value, where the excited pairs of dynamical fermions screen the static charges creating two isolated color singlets (also called mesons \cite{U1Zoller}). In Fig.~\ref{fig1}(b) the static potential for $Q/g=1$ and different values of $m/g$ shows that the smaller $m/g$, the easier it is to break the string, i.e., smaller values for $L_{c}$. The accuracy of the Gaussian variational method can be validated in comparison to MPS with relative error bounded by $0.7\%$ even for $L$ close to $L_{c}$.

In Fig.~\ref{fig1}(c) we consider noninteger values of $Q/g$ for $m/g=1$. In this case the static charges cannot be completely screened via production of pairs, which can only screen their integer part. This fact gives rise to the appearance of several string breaking processes with a remaining flux connecting the charges. In this case the relative error between the two ans{\"a}tze is bounded by $0.4\%$ in comparison with MPS results in Ref.~\cite{VQED}.

The reason why the Gaussian ansatz turns out to describe the model so accurately, can be understood as follows. In the limit $x\to \infty $, the Hamiltonian \eqref{H} becomes quadratic and then it can be exactly solved by the Gaussian ansatz. On the other hand, for $x\to 0$, the hopping term vanishes and the ground state is given by the bare vacuum \eqref{eq:SC_U1}. In fact, this state turns out to be a Gaussian state as well and therefore both limits are accurately captured by the Gaussian ansatz. This analysis together with the results for the charge and flux distribution profiles shows that the Gaussian ansatz accurately captures the equilibrium properties of the U(1) Schwinger model. In the following sections we return to our original parameter convention fixing $\varepsilon=1$ as unit of energy similar to the discrete models considered in Refs.~\cite{U1Zoller,kuhn}.

\begin{figure*}[tbp]
\begin{center}
\includegraphics[width=0.99\linewidth]{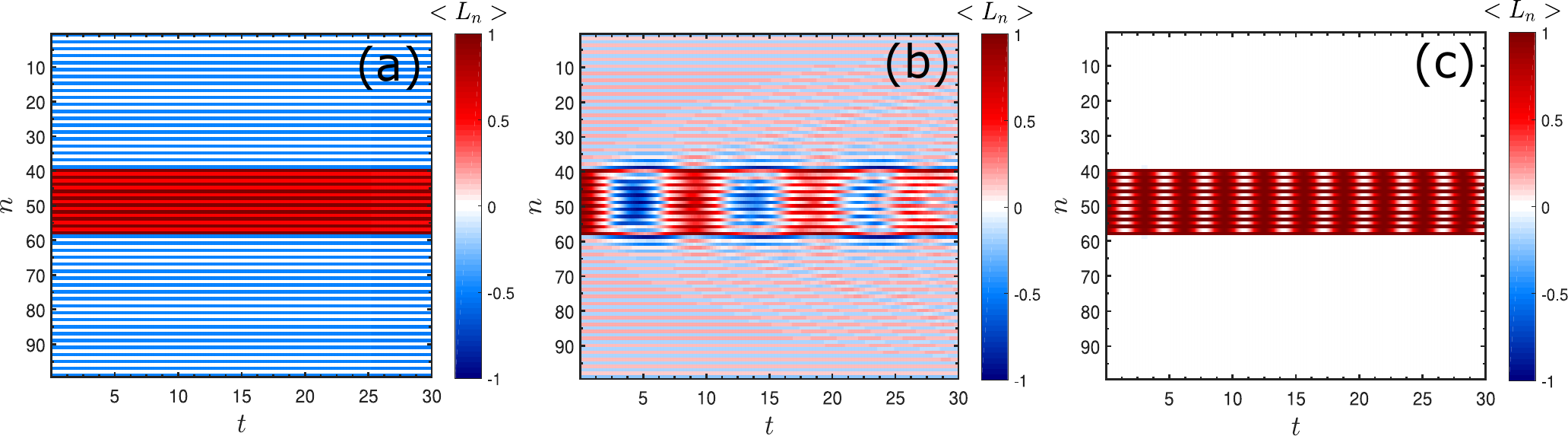}
\end{center}
\caption{Evolution of the electric flux distribution $\avg{L_n}$ for an initial static string of length $L/a=19$ and $Q/g=1$ imposed on top of the interacting vacuum. (a) The non-interacting case $m=g=0$. (b) The appearance of string and anti-string configurations for $m=0.1$, $g=1$. (c) The strong coupling limit $m=3$, $g=3.5$.}
\label{fig2}
\end{figure*}

\subsection{Real-time dynamics \label{sec:realU1}}
In this section we probe the validity of the Gaussian ansatz to describe dynamical properties of the Schwinger model by solving Eq.~\eqref{DynRe} for certain initial conditions. In particular we are interested in the dynamics of string breaking. We consider two different scenarios: (i) an initial dynamical string configuration whose ends at $n_{1}$ and $n_{2}$ can freely propagate, which we call the free string, given by the non-Gaussian state $\Theta ^{\dagger }\phi _{n_{1}}^{\dagger }\phi _{n_{2}}\left\vert \text{GS} \right\rangle \left\vert 0\right\rangle _{\text{gauge}}$ for odd distance; (ii) a string with two static charges on its ends, referred to as a static string. Moreover, time is measured in units of $1/\varepsilon $.

For the static string created on top of the interacting vacuum with $m,g\neq 0$, one expects that the string will dynamically break for the length $L/a=19$ which exceeds the critical distance. In Fig.~\ref{fig2}, the real-time evolution of the electric flux configuration in different parameter regimes for a static string is shown.

\begin{figure*}[tbp]
\begin{center}
\includegraphics[width=0.99\linewidth]{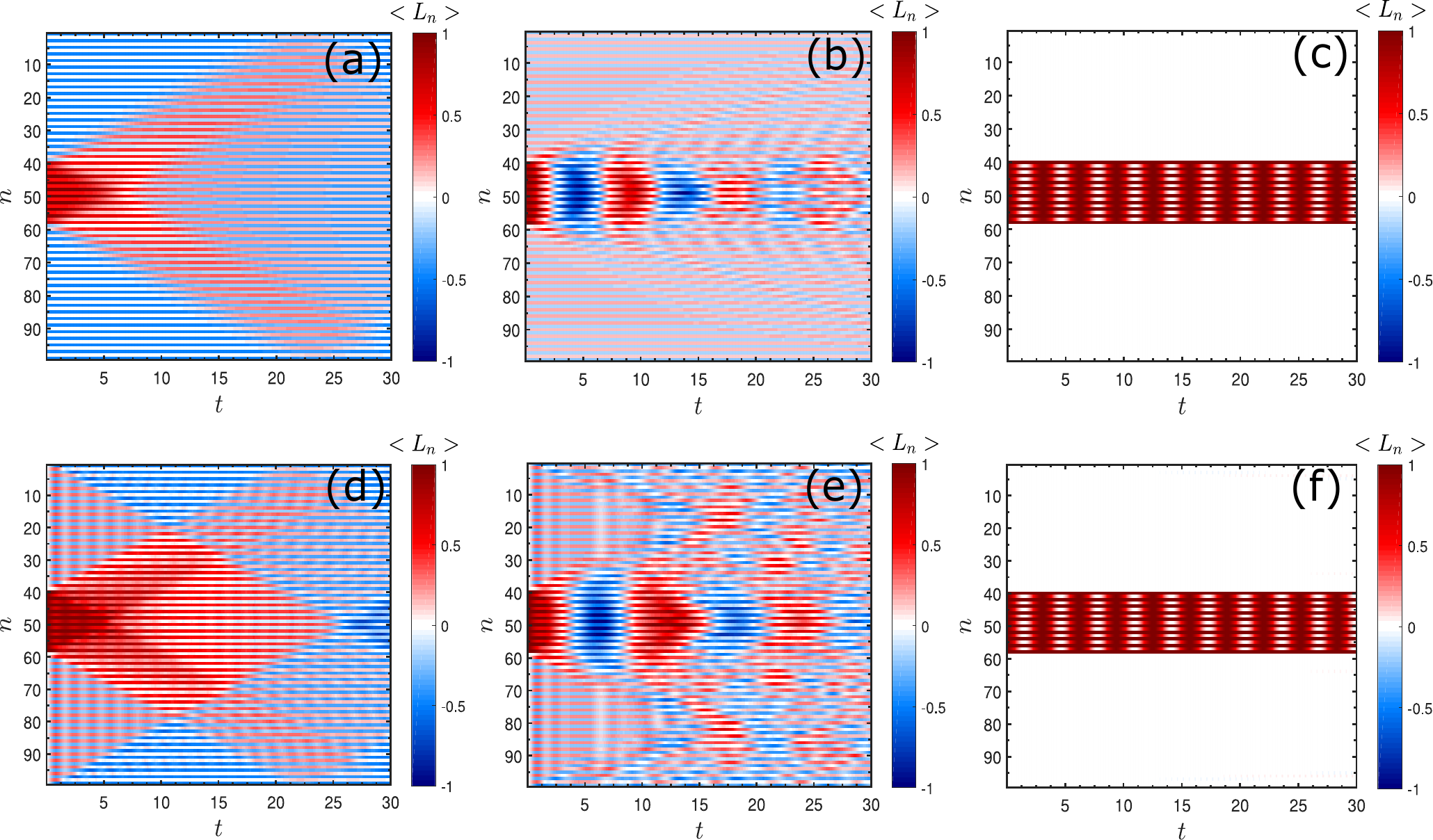}
\end{center}
\caption{Evolution of the electric flux distribution $\avg{L_n}$ for an initial free string of length $L/a=19$ and $Q/g=1$, the columns corresponds to $m=g=0$ (first column), $m=0.1$, $g=1$ (second column), and $m=3, g=3.5 $ (third column). Panels (a)-(c) show the evolution of a string imposed on top of the interacting vacuum, where no waves are coming from the boundaries. Panels (d)-(f) show the evolution of the string on imposed on top of the bare vacuum, leading to waves propagating from the boundaries and destructively interfering with the propagating string. The parameter regime $m=g=0$ used in panels (a) and (d) can be exactly solved with the Gaussian ansatz. Moreover, panels (b) and (e) show the Schwinger mechanism.}
\label{fig3}
\end{figure*}

In Fig.~\ref{fig2}(a) we show the electric flux for the noninteracting massless case $m=g=0$.  This case can be analytically solved with the Gaussian ansatz where the initial state is the interacting vacuum with half filling i.e., $\left\langle \phi _{n}^{\dagger }\phi_{n}\right\rangle =1/2$ for all sites $n$. The calculation shows that a line of electric flux connecting the static charges does not increase the energy of the system and therefore the string state is stationary, as can be seen in the figure.

In the intermediate regime, shown in Fig.~\ref{fig2}(b), a more intriguing feature emerges. We observe the formation of anti-string configurations, i.e., a string with opposite orientation of charges at its ends in comparison to the original string, in the center where the production of particle-antiparticle pairs takes place~\cite{U1Zoller}. This effect is called Schwinger mechanism where the creation of these pairs in a uniform electric field is viewed as the quantum process in which virtual pairs can be separated to become real pairs once they gain the binding energy of twice the rest mass energy \cite{DynStrBr1,pairprod}. We will further study this phenomenon in the following section.

Fig.~\ref{fig2}(c) shows the result for the strong coupling regime $m=3$, $g=3.5$. In this case, the interacting vacuum in the outer region is stable due to the high cost in energy of creating pairs out of the vacuum. Nevertheless an oscillation between a string and a broken string is realized in the center.

For the free string whose ends can freely propagate, we consider the string created on top of the interacting vacuum and the bare vacuum, i.e., the ground state of $\sum_n{(-1)^n\phi^{\dagger}_n\phi_n}$. The result for the bare vacuum is used to qualitatively compare our results with those obtained from the DMRG calculations in Ref.~\cite{U1Zoller}. We note that in Ref.~\cite{U1Zoller}, the gauge field was not eliminated and an effective Quantum Link model with spin-1 operators on the link was considered. Moreover, in order to qualitatively compare to the existing results, we plot our results for the electric field within the range $\abs{\avg{L_n}}\leq 1$.

The real-time evolution of the electric flux configuration for the initial free string is shown in Fig.~\ref{fig3}. For the noninteracting massless Schwinger case, which is exactly solvable with a Gaussian state, fermionic excitations can be created out of the vacuum at no energy cost which gives rise to large fluctuations of the electric field. Notice the difference in the evolution of the electric field in Figs.~\ref{fig3}(a) and \ref{fig3}(d) for a string created on top of the interacting and the bare vacuum, respectively. In the former case, Fig.~\ref{fig3}(a) shows that only the initial string propagates on the lattice. For the string imposed on top of the bare vacuum, the evolution corresponds to a global quench, which results in an interference of two wave fronts in the bulk as can be seen in Fig.~\ref{fig3}(d). The difference of our results in Fig.~\ref{fig3}(d) with respect to a similar study with a spin-1 Quantum Link model in Ref.~\cite{U1Zoller} can be explained due to the large fluctuations of the electric field, which lead to a maximal value $\max_{n,t}{\left\vert \left\langle L_{n}\right\rangle (t)\right\vert }=1.35>1$ at $t\approx 1$ during the evolution. As a result, the quantum link model in this parameter regime with $g=0$ is not equivalent to the Schwinger model considered in this paper \cite{pairprod}.

In Figs.~\ref{fig3}(b) and \ref{fig3}(e), we choose $m=0.1$, $g=1$. In this intermediate regime the string created on top of the interacting vacuum is breaking due to the creation of particle-antiparticle pairs, as Fig.~\ref{fig3}(b) reveals. At the beginning, string/anti-string configurations emerge alternatingly and finally disappear at a later time for which a steady state is reached. Fig.~\ref{fig3}(e) clearly shows that for a string imposed on top of the bare vacuum, we again observe an interference between the electric field wave fronts coming from the center and the boundary due to the quench dynamics.

In the regime of strong coupling and large mass, $m=3$, $g=3.5$, the Hamiltonian \eqref{H} is dominated by the mass and long-range interaction terms rendering the dynamical fermions outside the string region essentially static. As a result, the bare and interacting vacua are very similar. Thus, as shown in Figs.~\ref{fig3}(c) and \ref{fig3}(f), the time evolution of the electric field for a string imposed on the interacting vacuum is almost identical to the one imposed on the bare vacuum. In both cases the string is not completely broken. Because of the small fluctuations of the electric field in this regime, our results from Figs.~%
\ref{fig2}(c), \ref{fig3}(c) and \ref{fig3}(f) are in very good agreement with those obtained with a Quantum Link model in Ref.~\cite{U1Zoller}.

In general, we observe that the Gaussian ansatz captures the relevant features of the static and dynamical aspects of string breaking and correctly describes the production and propagation of particle-antiparticle pairs.

\section{SU(2) gauge theory \label{sec:SU2}}
In the previous section we demonstrated the suitability of the ansatz to  study Abelian gauge theories. However the Gaussian variational method is also adequate for studying non-Abelian lattice gauge models. To illustrate that,  we investigate an SU(2) LGT, which exhibits a richer Hilbert space structure compared to the Schwinger model. As before, we will consider the decoupled Hamiltonian \eqref{H} that only contains the fermionic matter fields. This allows us to correctly estimate vacuum energies in the absence of static charges via the variational ansatz $\ket{\psi}=\Theta^{\dagger}\ket{\text{GS}} \left\vert 0\right\rangle _{\text{gauge}}$. The situation is different if we consider static charges. Unlike the U(1) case, now they are noncommuting operators. As a result the non-trivial color entanglement between the external charges and dynamical fermions prevents the description with a Gaussian state $\ket{\text{GS}}$ even after applying the transformation $\Theta$.

To overcome this difficulty, we introduce two additional non-Gaussian unitary transformations, $V_{1}$ and $V_{2}$, which efficiently disentangle the static and the dynamical degrees of freedom. In the new frame, the static charges and the dynamical fermions are decoupled and the former appear as classical variables in the rotated Hamiltonian. While a transformation $V_1$ decoupling a single static charge from the dynamical fermions has been already used in the context of the Kondo model \cite{TaoKondo}, here we generalize this approach and introduce a new transformation $V_2$ to
decouple the second static charge. Moreover, we also introduce the correlation function between the static charges and between static charges and dynamical fermions. Using the variational ansatz $\ket{\psi}=\Theta^{\dagger}V_1^{\dagger}V_2^{\dagger}\ket{\text{GS}}\left\vert 0\right\rangle _{\text{gauge}}$, we analyze again ground state properties in the presence of two static charges as well as real-time dynamics.

\subsection{Decoupling the static charges \label{sec:DecVV}}
To study the string-breaking phenomenon we insert a pair of static charges at two sites $n_{1}$ and $n_{2}$. This corresponds to the static charge distribution
\begin{equation}
q_{n}^{a }=\frac{1}{2}(\delta _{n,n_{1}}\sigma _{1}^{a }+\delta
_{n,n_{2}}\sigma _{2}^{a }).
\end{equation}%

When the string is present, the pair of static charges forms the spin singlet state $\sum_{\alpha=r,g }{\Theta ^{\dagger }\phi _{n_{1}}^{\dagger,\alpha}\phi _{n_{2}}^{\dagger,-\alpha}}\left\vert \text{GS}\right\rangle\ket{0}_{\text{gauge}} $, and when the string is broken, each static charge forms a singlet state with the
surrounding dynamical fermions. As a result, neither of them is a Gaussian state.

However, if we are able to decouple the static charges $\bm{q}_{n}$ from the dynamical fermions with a unitary transformation $V_2V_1$, the transformed Hamiltonian (conditioned on the spin state of static charges) only contains  operators acting on the dynamical fermions which can be studied with the Gaussian state approach. In fact, this decoupling is possible if the Hamiltonian has certain parity symmetries.

The problem we are trying to tackle here resembles a two-impurity problem described by the Hamiltonian (\ref{H}), where the impurities are two static charges described by the Pauli matrices $\sigma ^{\alpha }$. The Hamiltonian $H_{\Theta }$ has the parity symmetry
\begin{equation}
\lbrack P_{1},H_{\Theta }]=0,
\end{equation}%
where the operator $P_{1}=\sigma _{1}^{z}\sigma _{2}^{z}P_{z}$ and $P_z$ is defined to be
\begin{equation}
P_{z}=\exp \Big[i\frac{\pi }{2}\sum_{n}{\bm{\phi}_{n}^{\dagger }(\sigma ^{z}+%
\mathds{1})\bm{\phi} _{n}}\Big].
\end{equation}%
This $\mathbb{Z}_2$ symmetry corresponds to the rotational invariance of the entire system along the $z$-direction by $\pi$.

Similar to the single-impurity Kondo model~\cite{TaoKondo}, we can construct the unitary transformation
\begin{equation}
V_{1}=\frac{1}{\sqrt{2}}(1-i\sigma _{1}^{y}\sigma _{2}^{z}P_{z}),
\end{equation}%
which transforms $P_{1}$ into the operator $\sigma _{1}^{x}$ of the first impurity, $V_{1}P_{1}V_{1}^{\dagger }=\sigma _{1}^{x}$. Since $P_{1}$ is a symmetry of
the Hamiltonian, $\sigma _{1}^{x}$ is conserved in the new frame, i.e.,
\begin{equation}
\lbrack \sigma _{1}^{x},H_{1}]=0,
\end{equation}%
and can be considered as a \textquotedblleft classical\textquotedblright\ variable, where $H_{1}=V_{1}H_{\Theta }V_{1}^{\dagger }$. The explicit form of $H_{1}$
is
\begin{align}
H_{1}& =\varepsilon\sum_{n}\big(\bm{\phi}_{n}^{\dagger }\bm{\phi}_{n+1}+\textrm{H.c.}\big) \\
&+m \sum_{n}(-1)^{n}\bm{\phi}_{n}^{\dagger }\bm{\phi}_{n}+\frac{g^2}{2}H_{e},
\end{align}%
where the electric term reads
\begin{align}
H_{e} =&\frac{1}{4}\sum_{a }\sum_{kp}{V}_{kp}\bm{\phi}_{k}^{\dagger
}\tau ^{a }\bm{\phi}_{k}\bm{\phi}_{p}^{\dagger }\tau ^{a }\bm{\phi}%
_{p}\notag \\
&+\frac{1}{2}\sum_{k}{V}_{kn_{1}}(\sigma _{1}^{x}\bm{\phi}_{k}^{\dagger
}\tau ^{x}\bm{\phi}_{k}-i\sigma _{2}^{z}P_{z}\bm{\phi}_{k}^{\dagger }\tau
^{y}\bm{\phi}_{k}\notag \\
& +\sigma _{1}^{x}\sigma _{2}^{z}P_{z}\bm{\phi}_{k}^{\dagger }\tau ^{z}%
\bm{\phi}_{k}) +\frac{1}{2}\sum_{k,a}{V}_{kn_{2}}\sigma _{2}^{a }%
\bm{\phi}_{k}^{\dagger }\tau ^{a}\bm{\phi}_{k}\notag \\
&+\frac{1}{2}{V}%
_{n_{1}n_{2}}(\sigma _{1}^{x}\sigma _{2}^{x}-\sigma _{2}^{x}P_{z}+\sigma
_{1}^{x}P_{z}).
\end{align}
To decouple the second static charge located at $n_{2}$, we notice that the Hamiltonian $H_{1}$ is rotationally invariant along the $x$-direction for an even number $\mathcal{N}=\sum_{n}{\bm{\phi}_{n}^{\dagger }\bm{\phi}_{n}}$ of dynamical fermions. Thus, $P_{2}=\sigma _{2}^{x}P_{x}$ is the parity symmetry of the Hamiltonian $H_{1}$, where%
\begin{equation}
P_{x}=\exp \Bigl[i\frac{\pi }{2}\sum_{n}{\bm{\phi}_{n}^{\dagger }(\sigma ^{x}+\mathds{1})\bm{\phi}_{n}}\Bigr].
\end{equation}

A second unitary transformation%
\begin{equation}
V_{2}=\frac{1}{\sqrt{2}}(1-i\sigma _{2}^{y}P_{x})
\end{equation}%
allows us to rotate the parity operator $P_{2}=\sigma _{2}^{x}P_{x}$ for the second charge, as $V_{2}P_{2}V_{2}^{\dagger }=-\sigma_{2}^{z}$. Thus, in the new frame, $\sigma _{2}^{z}$ commutes with the Hamiltonian $H_{2}=V_{2}H_{1}V_{2}^{\dagger }$. The final form of the transformed Hamiltonian is given by
\begin{widetext}
\begin{align}
\begin{aligned}
H_2(\sigma_1^x,\sigma_2^z)&=\varepsilon\sum_n{\big(\bm{\phi}_n^{\dagger}\bm{\phi}_{n+1}+\textrm{H.c.}\big)}+m\sum_n{(-1)^n\bm{\phi}_n^{\dagger}\bm{\phi}_n}+\frac{g^2}{2}\Big[\sum_{a}{\sum_{n,m}{Q_n^{a}V_{n,m}Q_m^{a}}}\\
&+\frac{1}{2}V_{n_1,n_2}\Big(-\sigma_1^x\sigma_2^zP_x+i^{\mathcal{N}}\sigma_2^zP_y+\sigma_1^xP_z \Big)+\sum_{m}{V_{n_1,m}\Big( \sigma_1^xQ_m^x-i\sigma_2^zP_zQ_m^y+\sigma_1^x\sigma_2^zP_zQ_m^z\Big)}\\ & +\sum_{m}{V_{n_2,m}\Big( -\sigma_2^zP_xQ_m^x-iP_xQ_m^y+\sigma_2^zQ_m^z\Big)}\Big]
\end{aligned}
\label{HSU2dec}
\end{align}
\end{widetext}
which only depends on the two commutating operators $\sigma_1^x,\sigma_2^z$ for the static charges, where the operator $P_{y}$ is defined as
\begin{equation}
P_{y}=\exp \Big[i\frac{\pi }{2}\sum_{n}{\bm{\phi}_{n}^{\dagger }(\sigma ^{y}+%
\mathds{1})\bm{\phi}_{n}}\Big],
\end{equation}%
and the relation%
\begin{equation}
P_{x}P_{z}=P_{z}P_{x}e^{i\pi\mathcal{N}}=P_{z}P_{x}=i^{\mathcal{N}}P_{y}
\end{equation}%
has been used assuming an even number of dynamical fermions.

A remarkable feature of this Hamiltonian is that the operators for static charges become classical variables. Indeed, a general state in the rotated frame can be written as $\left\vert \Psi \right\rangle =\left\vert \Psi _{2}\right\rangle \left\vert s_{1}\right\rangle \left\vert s_{2}\right\rangle $, where $\left\vert s_{1}\right\rangle $ and $\left\vert s_{2}\right\rangle $ are eigenstates of $\sigma _{1}^{x}$ and $\sigma _{2}^{z}$, respectively. The evolution of the state $\left\vert \Psi _{2}\right\rangle $ is
governed by the Hamiltonian $H_{2}(s_{1},s_{2})$ of dynamical fermions and will be approximated by the Gaussian state $\ket{\text{GS}}$.

To show the decoupling procedure and the corresponding symmetries in a compact way, we can rewrite the symmetry operators as
\begin{equation}
P_{1}=-i^{\mathcal{N}}e^{i\pi\mathcal{Q}^{z}}
\end{equation}
and
\begin{equation}
\bar{P}_{2}=V_{1}^{\dagger }P_{2}V_{1}=(-1)^{\mathcal{N}}e^{i\pi\mathcal{Q}^{y}}
\end{equation}
in the original frame before applying $V_{1}$ and $V_{2}$, where $\mathcal{Q}^{z}=\sum_{n}{\mathcal{Q}_{n}^{z}}$ and $\mathcal{Q}^{y}=\sum_{n}{\mathcal{Q}_{n}^{y}}$ are the total (dynamical plus external) SU(2)-charge operators along the $z$ and the $y$ directions. Apart from the prefactor determined by the fixed dynamical fermion number, these two symmetries correspond to the rotational invariance of the original Hamiltonian $H_{\Theta }$ along the $z$ and the $y$ directions by angle $\pi$. The analysis above implies that a single unitary transformation $V_{2}V_{1}$ can be applied directly to transform both symmetry operators to $\sigma_{1}^x$ and $-\sigma_{2}^z$ of static charges, respectively. In the new frame, $\sigma_{1}^x$ and $\sigma_{2}^z$ commute with the Hamiltonian $H_2$, and, thus, become the classical variables $s_1$ and $s_2$. Correspondingly the symmetry operators $P_{1}$ and $\bar{P}_{2}$ take the values $s_1$ and $-s_2$, respectively. As will be shown in Sec.~\ref{Sp}, the relation of $s_{1,2}$,  $\mathcal{N}$, and $\mathcal{Q}^{z,y}$ provides us a clear picture of the ground state configuration from the symmetry analysis.

We emphasize that $V_{1}$ and $V_{2}$ are two non-Gaussian unitary transformations entangling static charges and dynamical fermions. This entanglement induced by $V_{1}$ and $V_{2}$ can be seen from the variational state
\begin{align}
\begin{aligned}
& \frac{1}{4\sqrt{2}}\Theta ^{\dagger }\{(\ket{\uparrow}_{z}+s_{1}%
\ket{\downarrow}_{z})\big[(1+s_{2})\ket{\uparrow}_{z}+(1-s_{2})%
\ket{\downarrow}_{z}\big]  \\
& +s_{2}(s_{1}\ket{\uparrow}_{z}-\ket{\downarrow}_{z})\big[(1+s_{2})%
\ket{\uparrow}_{z}+(1-s_{2})\ket{\downarrow}_{z}\big]P_{z}  \\
& -s_{2}(\ket{\uparrow}_{z}+s_{1}\ket{\downarrow}_{z})\big[(1-s_{2})%
\ket{\uparrow}_{z}+(1+s_{2})\ket{\downarrow}_{z}\big]P_{x} \\
& +(s_{1}\ket{\uparrow}_{z}-\ket{\downarrow}_{z})\big[(1-s_{2})\ket{\uparrow}%
_{z}+(1+s_{2})\ket{\downarrow}_{z}\big]i^{\mathcal{N}}P_{y}\}\\
&\times\ket{\text{GS}}\left\vert 0\right\rangle _{\text{gauge}}
\end{aligned}
\label{ansatz}
\end{align}
in the original frame, which correctly captures the physics for the appropriate choice of the parameters $s_1,s_2\in\{-1,+1\}$ and the Gaussian state. Note that if the Gaussian state is taken to be the Dirac sea (Eq.~\eqref{eq:SC_SU2}), it is a common eigenstate of all operators $P_c$ with the same eigenvalue (either $+1$ or $-1$) and Eq.~\eqref{ansatz} becomes the singlet state between static charges for $s_1=s_2=-1$, while it becomes a triplet state for any other choice of $s_1$ and $s_2$. However, unlike the Dirac sea, a general Gaussian state $\ket{\text{GS}}$ does not preserve the rotational symmetry of the Hamiltonian, and gives rise to entanglement between static and dynamical fermions. We will explore this entanglement structure in greater detail in our studies of the ground state and the real-time dynamics.

To explicitly characterize the entanglement between static charges and dynamical fermions, we introduce two gauge invariant correlation functions
\begin{equation}
C_{2}(n_{1},n_{2})=\sum_{a,b }{\left\langle q_{n_{1}}^{a }%
\Big(U_{n_{1}}^{\text{Adj.},\dagger }\cdots U_{n_{2}-1}^{\text{Adj.},\dagger
}\Big)_{a,b}q_{n_{2}}^{b}\right\rangle }
\end{equation}%
between the static charges, and
\begin{equation}
C_{\text{dyn}}(n_1,n)=\sum_{a,b }{\left\langle {q}^{a}_{n_{1}}\Big(
	U_{n_{1}}^{\text{Adj.},\dagger }\cdots U_{n-1}^{\text{Adj.},\dagger
	}\Big)_{a,b}{Q}^b_{n}\right\rangle }
\end{equation}
between the static charge at $n_1$ and the dynamical fermion at $n$. Note that in the gauge-field-free frame, these correlation functions become
\begin{equation}
C_{2}(n_{1},n_{2})=\sum_{a }{\left\langle q_{n_{1}}^{a }%
q_{n_{2}}^{a }\right\rangle },
\end{equation}
and
\begin{equation}
C_{\text{dyn}}(n_1,n)=\sum_{a }{\left\langle q_{n_{1}}^{a }%
Q_{n}^{a }\right\rangle }.
\end{equation}

In the limit where the string dominates (e.g. in the initial state, when the string is superimposed on the bare vacuum), the two static charges form a singlet state through the flux string connecting them, hence $C_{2}(n_{1},n_{2})$ approaches $-3/4$. In the opposite regime, when the string is broken, a color singlet is formed between each static charge and the dynamical fermions, and the correlation $C_{2}(n_{1},n_{2})$ vanishes. Hence, this correlation function indicates the occurrence of string breaking
\begin{equation}
C_{2}(n_{1},n_{2}) \to
\begin{cases}
-3/4 & \text{ in the string regime } \\
0 & \text{ in the broken-string regime }
\end{cases}
\end{equation}
In the latter case, the entanglement between the static charge at $n_1$ and the screening ``cloud'' of dynamical fermions can be characterized by  $C_{\text{dyn}}(n_1,n)$. Furthermore, the correlation functions enable us to give a precise description of entanglement structure between the static charges and the dynamical fermions for in and out-of-equilibrium dynamics.

Analogously to Sec. \ref{sec:U1} for the U(1) gauge group, we make use of the evolution Eqs.~\eqref{DynIm} and \eqref{DynRe} for $\Gamma$ to study the string-breaking phenomenon in the ground state and its real-time dynamics. The effective mean-field Hamiltonian appearing in the Eqs.~\eqref{DynIm} and \eqref{DynRe} is derived in Appendix \ref{sec:SU2eff}. Unlike the Abelian case, the Hamiltonian $H_{2}(s_{1},s_{2})$ contains exponential functions of the creation and annihilation operators, hence the derivation of the mean-field Hamiltonian follows a much more sophisticated procedure~\cite{Tao}.

\subsection{Static properties\label{Sp}}
Let us first apply the Gaussian ansatz in combination with the unitary transformations $\Theta $, $V_{1}$ and $V_{2}$ to study the static aspects of the string-breaking phenomenon in the ground state of the SU(2) LGT. Additionally, to benchmark the Gaussian variational ansatz, we also perform MPS calculations (see Appendix \ref{app:mps} for details) and compare the results of both simulations. We observe that with our choice of parameters in both optimization algorithms, numerical artifacts are negligible for the effects we want to observe. Note that again we measure energy in units of $\varepsilon=1$.

First of all, we compare the values for the ground state energies in the absence of static charges i.e., the vacuum energies $E_{\text{vac}}/\varepsilon$. In Table~\ref{tab:Evac}, considering MPS as the reference result, we show the relative difference
\begin{equation}
\Delta=\frac{E_{\text{vac}}^{\text{Gauss}}-E_{\text{vac}}^{\text{MPS}}}{\left| E_{\text{vac}}^{\text{MPS}} \right|},
\end{equation}
between Gaussian and MPS ans{\"a}tze for  $E_{\text{vac}}$. As expected, the MPS simulation yields lower values for the vacuum energy in all cases but even in the strong coupling regime (with $m=0.5$ and $g=3$) the relative error is bounded below $2.1\%$. Moreover we observe that the larger the coupling $g$ becomes with respect to the mass $m$ and the hopping $\varepsilon$, the more relevant the (non-quadratic) electric energy term $H_e$ becomes, and, thus, resulting in larger errors.

\newcolumntype{M}[1]{>{\centering\arraybackslash}m{#1}}
\newcolumntype{N}{@{}m{0pt}@{}}
\begin{table}[!htb]
\centering
\begin{tabular}{|M{1cm}|M{1.1cm}|M{1.5cm}|M{1.5cm}|M{1cm}|N}
\hline
\multicolumn{5}{|c|}{\text{E$_{\text{vac}}/\varepsilon$}} \\ \hline
$m$ & $g$ & Gaussian & MPS & $\Delta$ ($\%$) &\\[7pt] \hline
1 & 1 & -64.2559 & -64.2760 & 0.03 \\ \hline
0.5 & 1 & -51.9578  & -52.0270 & 0.13 \\ \hline
0.75 & 1.5 & -54.6641 &  -54.7920  & 0.23 \\ \hline
0.5 & 1.5 & -48.2457 &-48.4827  & 0.49 \\ \hline
0.5 & 2 & -43.9929 & -44.4673 & 1.07 \\ \hline
0.5 & 3 & -36.1139 & -36.8746 & 2.06 \\ \hline
\end{tabular}
\caption{Comparison between the vacuum energies $E_{\text{vac}}/\varepsilon$ obtained from the Gaussian ansatz and the MPS calculation.}
\label{tab:Evac}
\end{table}

\begin{figure*}[tbp]
\begin{center}
\includegraphics[width=0.99\linewidth]{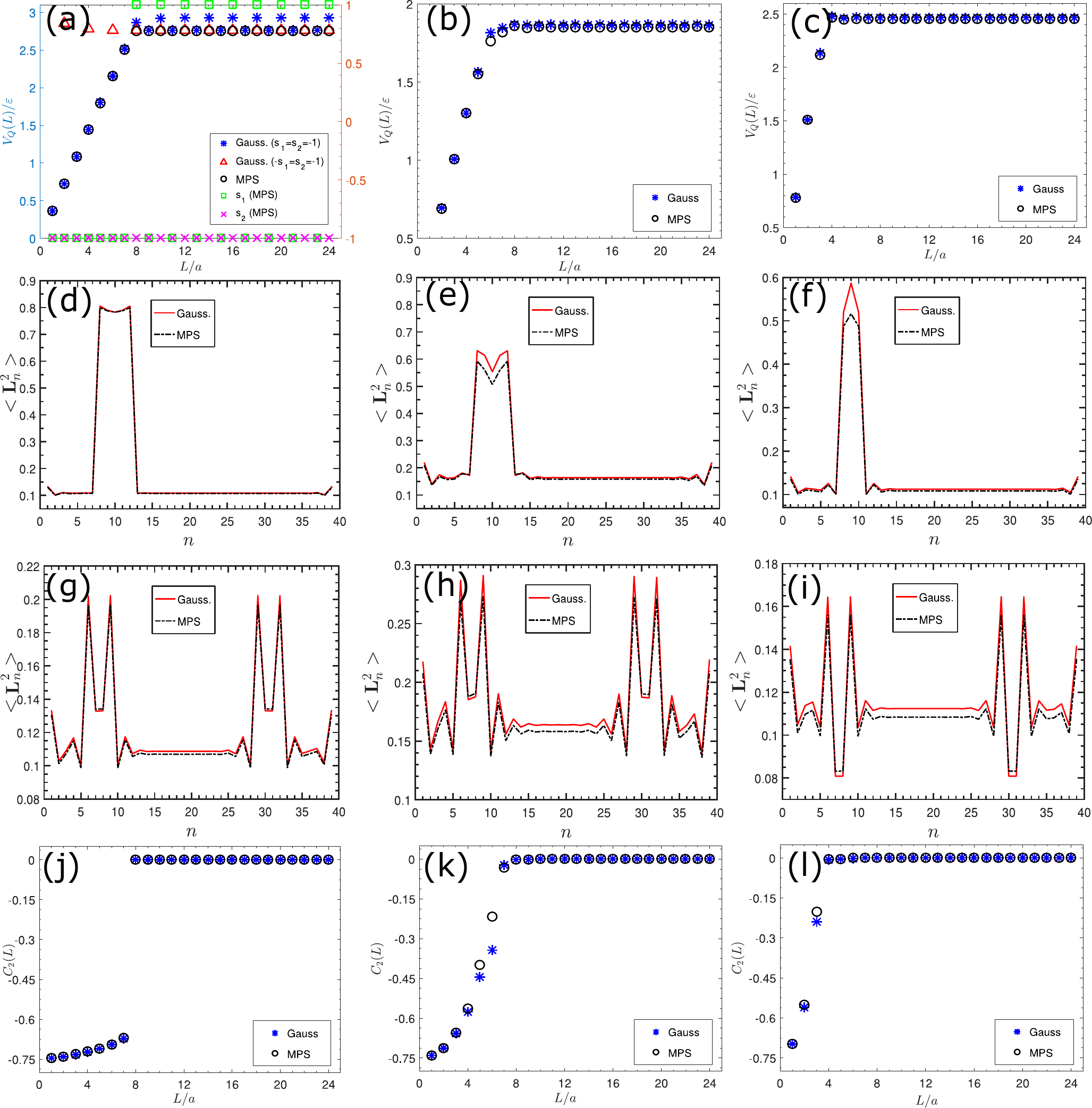}
\end{center}
\caption{Comparison between the equilibrium properties of the SU(2) LGT from the non-Gaussian ansatz and the MPS simulations. The columns correspond to parameters $m=1, g=1$ (first column), $m=0.5, g=1$ (second column), and  $m=0.75, g=1.5$ (third column). Panels (a)-(c) in the first row show the static potential $V_Q(L)/\varepsilon$. Additionally, in panel (a) we compare the results for different choices of $s_1$ and $s_2$ for the non-Gaussian ansatz in Eq.~\eqref{ansatz} and verify the result computing the expectation values for the parity operators $P_1$ and $P_2$ in the global ground state determined via a MPS simulation. Panels (d)-(f) show the color-electric flux profiles at various separations $L/a$ between the external charges yielding a string ground state. Analogously, panels (g)-(i) show the flux profiles for separations where the flux string is broken. Finally, the panels (j)-(l) in the last row show the correlation function $C_2(L)$ between static charges in the ground state as a function of the distance.}
\label{fig4}
\end{figure*}

After comparing the interacting vacuum energies, we proceed to study the system in the presence of two static charges. Due to the parity symmetries, we conclude that the Hamiltonian has four sectors labeled by the eigenvalues $s_1$ and $s_2$ of $P_1$ and $-P_2$. Fixing $s_1,s_2\in\{-1,+1\}$, we can compute the ground states in all of these four sectors with Eq.~\eqref{DynIm}. The global ground state is then the one with the lowest energy. In Fig.~\ref{fig4} we present our results for $m=1$, $g=1$ (Fig.~\ref{fig4}(a)),  $m=0.5$, $g=1$ (Fig.~\ref{fig4}(b)), and $m=0.75$, $g=1.5$ (Fig.~\ref{fig4}(c)).

In Fig.~\ref{fig4}(a) the blue asterisks correspond to the ground state energy in the sector $s_1=s_2=-1$. As expected, before the string breaks, the ground state energy (potential energy) grows linearly with the distance of the static charges. In the broken string case, the ground state energy in the sector $s_1=s_2=-1$ shows an oscillatory behavior between odd and even distances. Computing the ground states in all four sectors, we find that for even distances the global ground state in the broken string regime is in the sector $s_1=1,s_2=-1$, and has the same energy as for odd distances, as shown by the red triangles in Fig.~\ref{fig4}(a). The energy of the global ground state (potential energy) obtained from the non-Gaussian ansatz forms the expected plateau once the string has broken and quantitatively agrees with that from the MPS calculations (black circles). Computing the expectation value of $P_1$ and $-\bar{P}_{2}$ in the ground state obtained from the MPS calculations we observe the same values for $s_1$ and $s_2$ in the different regimes. As shown by the green squares and the pink crosses the MPS results confirm that the global ground state is in the sector $s_1=s_2=-1$ in the string and string-breaking cases for odd distances, while it is in the sector $s_1=1$ and $s_2=-1$ for even distances in the broken string state. To understand why $s_1$ takes different values for even and odd distances in the string-broken state, one can inspect the total SU(2) charges $\mathcal{Q}^a$ (including both static charges and dynamical fermions) and the total number of dynamical fermions for even and odd distances. It turns out that the ground state for odd distances is half-filled, and the total SU(2) charge is zero, which leads to $s_1=s_2=-1$ for the system size $N=40$. However, for even distances, the ground state of the system in the broken string case always prefers to add or reduce two dynamical fermions and as a result, $s_2$ is kept the same and $s_1$ changes the sign.

In Figs.~\ref{fig4}(b) and \ref{fig4}(c), the potential energy $V_Q(L)$, is plotted for $m=0.5$, $g=1$ and $m=0.75$, $g=1.5$. As  $m$ decreases, i.e., for relatively stronger interactions, the description given by the non-Gaussian ansatz still agrees with the MPS result very well, with a relative error smaller than $1.1\%$.

The presence or not of the string in the ground state can also be distinguished in the electric flux profiles. These are shown in the second and third rows of Fig.~\ref{fig4} for two sets of distances corresponding to the string and the string-breaking regimes, with the first static charge always located at $n_1=8$. In the string case, the two static charges are confined by the flux string. In the string-breaking regime, each of the static charges is completely screened by the surrounding dynamical fermions, and the flux string is broken. Because of the global SU(2) symmetry each group component contributes equally to $\left\langle \mathbf{L}_n^2\right\rangle $. The flux distribution is also in good agreement with the one obtained from the MPS calculations, as shown by the red solid curves.

In the last row of Fig.~\ref{fig4}, the correlation function $C_2(L)$ between the static charges as a function of the distance is shown. The breaking of the string takes place at some distance $L_c$. For much shorter distances ($L \ll L_c$) $C_2(L)$ is close to $-3/4$ indicating that the static charges are confined by the electric flux string and form a singlet state. In contrast, for much longer distances ($L \gg L_c$) $C_2(L)$ is essentially $0$, thus implying that the static charges are screened by the surrounding dynamical fermions and the string is broken. Close to the distance at which the breaking takes place, the correlation function calculated from both the non-Gaussian ansatz (blue asterisks) and the MPS results (black circles) displays a sudden change for $m=1$, $g=1$ (Fig.~\ref{fig4}(e)) and a smooth transition for $m=0.5$, $g=1$ and $m=0.75$, $g=1.5$ (Figs.~\ref{fig4}(f) and \ref{fig4}(g)).

\begin{figure}[tbp]
\includegraphics[width=0.99\linewidth]{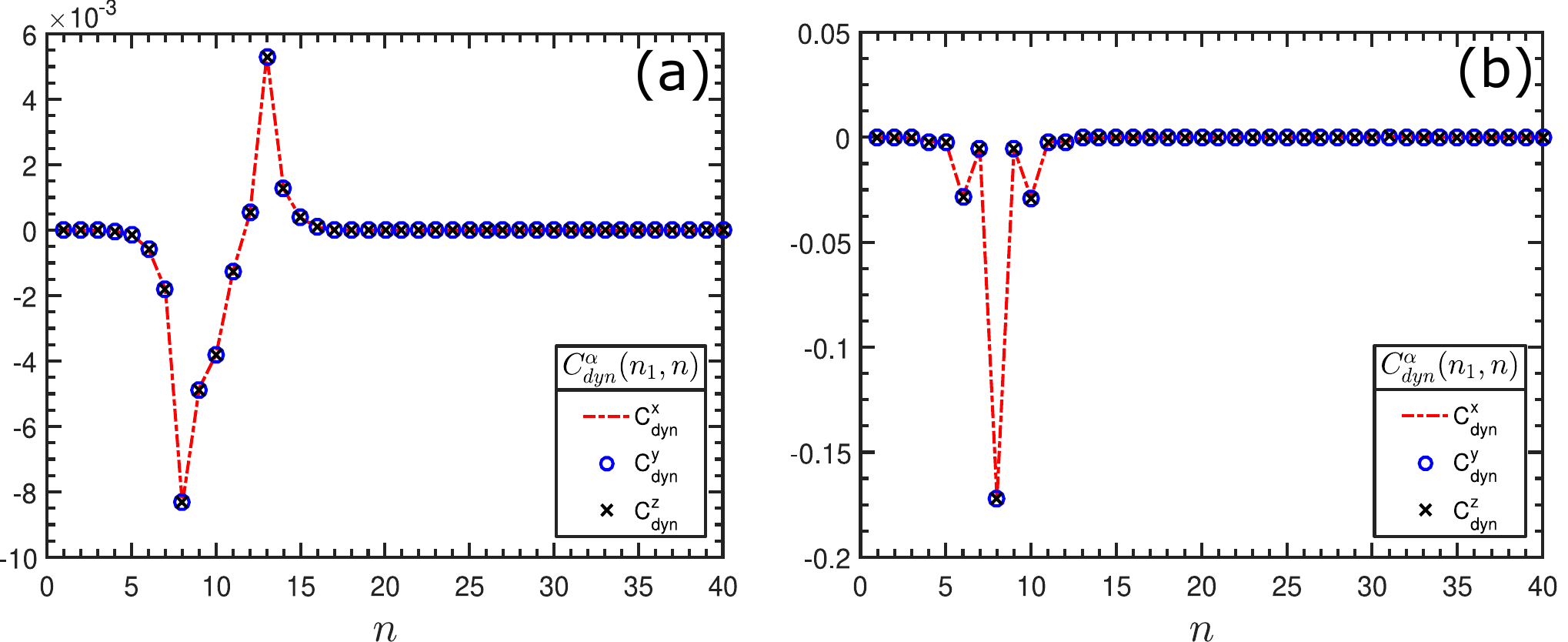}
\caption{Correlation function $C_{\text{dyn}}^{\alpha}(n_1,n)$ between the first static charge and dynamical fermions for $m=1$, $g=1$ and distances $L/a=5$ (a) and $L/a=23$ (b) between static charges.}
\label{fig5}
\end{figure}

The correlation function $C_{\text{dyn}}^{a}(n_1,n)$ between the first static charge and the dynamical fermions also displays different patterns in the string and string-breaking regimes. As shown in Fig.~\ref{fig5}, the static charges are hardly entangled with the dynamical fermions when the string is present, hence resulting in small values for $C_{\text{dyn}}^{a}(n_1,n)$. In contrast, $C_{\text{dyn}}^{a}(n_1,n)$ reveals that a symmetric cloud of dynamical fermions screens the static charges when the string is broken. In particular, due to the global SU(2) symmetry of the Hamiltonian each group component contributes equally, we see that $\sum_{n,a} C_{\text{dyn}}^{a}(n_1,n)$ is close to $-3/4$. Hence, the external charges are forming approximately a SU(2) singlet with the surrounding dynamical fermions.

Therefore, our results for the potential energy, the electric flux profiles, and the correlation functions provide a comprehensive description of the string-breaking phenomenon. In the string case, the two static charges connected by the electric flux string form a SU(2) singlet state, and are only weakly entangle with the half-filled dynamical fermions. As the distance increases, the flux string becomes longer and longer, which leads to the linear increase in the potential energy of static charges. At distances larger than a certain threshold value $L_c$, each static charge forms a color singlet with the dynamical fermions in the screening cloud, and the string is broken. For odd distances, the static charges are on sites of different staggered mass. This term favors double occupancy in one of them, and vacancy in the other. On the contrary, for even distances both static charges are located at sites favoring double occupancy with dynamical fermions (vacancy). Thus two fermions are annihilated (created) to screen the static charges. The screening results in a single dynamical fermion being localized on each of the static charges. In both cases the potential energy reaches a plateau indicating the occurrence of the string breaking.

All in all, the non-Gaussian ansatz accurately describes the string-breaking phenomenology in the (1+1)-dimensional SU(2) LGT, where the entanglement between the static charges and the dynamical fermions is entirely encoded by the transformations $V_1$ and $V_2$. It also allows us to effectively determine the relevant observables such as the potential energy, the electric flux profiles and the correlation functions.

\subsection{Real-time dynamics \label{sec:reSU2}}
In this subsection, we study the dynamical aspects of the string-breaking phenomenon using the non-Gaussian ansatz from Eq.~\eqref{ansatz}. Since the transformations $V_1$ and $V_2$ can only characterize the entanglement between static charges and dynamical fermions, but not the entanglement between different dynamical fermions, we restrict ourselves to study the dynamical features of static strings.

\begin{figure*}[tbp]
\begin{center}
\includegraphics[width=0.99\linewidth]{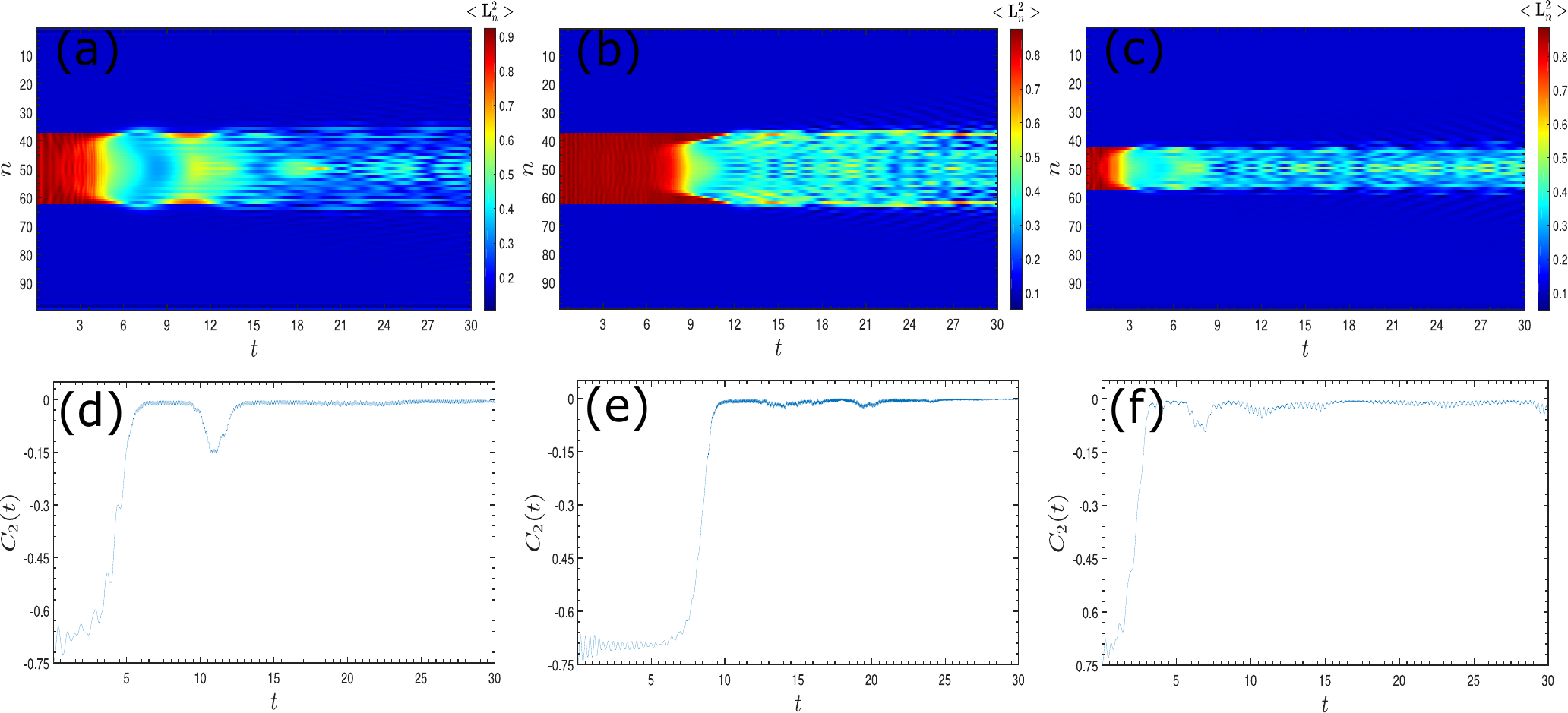}
\end{center}
\caption{Evolution of a color-flux string for an initial spin-singlet state between the static charges. Panels (a)-(c) show the evolution of the flux profile, panels (d)-(f) the evolution of the correlation function $C_2(t)$. The different columns correspond to the parameters  $m=0.5$, $g=1$, $L/a=25$ (first column), $m=0.75$,  $g=1.5$, $L/a=25$ (second column) and $m=0.75$, $g=1.5$, $L/a=15$ (third column).}
\label{fig6}
\end{figure*}

In Fig.~\ref{fig6}, the real-time evolution of the flux distribution $\left\langle \bm{L}_{n}^{2}\right\rangle (t)$ and the correlation function $C_{2}(t)$ are depicted for the static string imposed on top of the interacting vacuum for various combinations of $m$ and $g$.

Figs.~\ref{fig6}(a)-\ref{fig6}(c) shows that the interacting string is stable at short time scales, before it eventually breaks. Similar to the U(1) case, the time scale of the breaking process $t_{\text{SB}}$ depends on the length of the initial string, the mass $m$ and the coupling constant $g$ (recall we have fixed $\varepsilon=1$). Moreover, the string-breaking behavior is also captured by the correlation function $C_{2}(t)$ between the two static charges, as shown in the second row of Fig.~\ref{fig6}.

At initial times, the static charges connected by the flux string are in a singlet state, thus $C_{2}(0)\approx-3/4$. For short time scales, the dynamical fermions start to accumulate around the static charges and partially break the string, which corresponds to a damped oscillation in $C_{2}(t)$. Eventually, at $t_{\text{SB}}$ the static charges are completely screened by the surrounding dynamical fermions, the flux string is completely broken and $C_{2}(t)$ oscillates around $0$.

Comparing Figs.~\ref{fig6}(d) and \ref{fig6}(e) to Fig.~\ref{fig6}(f), we find that the larger the initial length $L$ is, the longer it takes for the string to break. This can be explained as follows. For larger distances $L$, the system requires a longer time to relax from the initial interacting string state to the string-breaking state by transporting a dynamical fermion between the two sites occupied by the static charges. Moreover, for the same distance between static charges, to decrease the mass $m$ and coupling constant $g$, effectively implies to increase the hopping parameter $\varepsilon$, which speeds up the screening process. Therefore, the string breaks faster for smaller values of $m$ and $g$, as can be seen in Figs.~\ref{fig6}(d) and \ref{fig6}(e).

\section{Conclusion and Outlook \label{sec:out}}
In this paper, we introduce a new family of variational ans{\"a}tze which is suitable to study lattice gauge models. Our method relies on three unitary transformations that rotate the original Hamiltonian \eqref{HU}. First, we derive $\Theta$ which decouples the matter and the gauge degrees of freedom. The resulting rotated Hamiltonian \eqref{H} is completely general and can be addressed with any numerical or analytical technique. For the U(1) LGT, the Gaussian ansatz can describe the interacting vacuum state and the in and out of equilibrium transition between the string and broken string cases. However, for the SU(2) LGT, while the ansatz is able to describe the interacting vacuum, it cannot be directly applied to study the string breaking in the presence of two static charges. Thus, we further introduce a non-Gaussian ansatz with two unitary transformations $V_{1}$ and $V_{2}$ that characterize the entanglement between static charges and dynamical fermions.

Using the Gaussian ansatz, we investigate the static and dynamical aspects of string breaking in the Schwinger model. First, we compute the interacting vacuum of the theory in the presence of two external charges and determine the static potential. For small distances between the external charges, we clearly see a linear increase in the static potential, thus indicating a flux string is present in the ground state. For larger distances beyond a critical one, it is energetically favorable to break the flux string resulting in a flattening of the potential. In general, we observe excellent agreement with previous tensor networks studies of the model~\cite{VQED} and we are able to precisely determine the regimes in which string breaking occurs. Second, we simulate the real-time dynamics of a flux string between two external charges. Computing the site resolved flux profiles we can also clearly distinguish between a string state and the breaking case in the out-of-equilibrium scenario. Even if we perform a global quench on the initial string state, we are able to simulate the dynamics with the Gaussian variational method.

This variational ansatz is not limited to the Abelian case and we can also explore static as well as dynamical properties for a SU(2) LGT in the presence of external charges. Since in this case the Gaussian ansatz cannot be directly applied to the rotated Hamiltonian \eqref{H}, we show how to overcome this limitation with the two additional unitary transformations $V_1$ and $V_2$. These two transformations, which decouple the static charges and the dynamical fermions, not only allow us to address the resulting Hamiltonian with fermionic Gaussian states, but also shed light on the entanglement structure between the dynamical fermions and the static charges. In the presence of a string we observe that the external charges are correlated among themselves. In contrast, in the breaking case correlations between the dynamical fermions and the static charges develop. Together with the static potential, the site resolved color-flux profiles and the correlation functions between charges, this allows us to give a comprehensive description of the string-breaking phenomenon. As for the Schwinger model, our results are in very good agreement with those obtained from MPS simulations. Furthermore, we also simulate the evolution of a color-flux string in real-time. Again, the flux profiles as well as the real-time evolution of the correlation function $C_2(t)$, clearly indicate whether the initial color flux string is breaking.

On the one hand, the unitary transformation $\Theta$ is not limited to the two specific cases we address and can be used with arbitrary SU($N$) gauge groups. More generally, it might be possible to derive a similar transformation for certain discrete gauge groups extensively used in the context of condensed matter physics. There is not yet a transformation to eliminate the gauge field in higher dimensions, i.e., 2+1 and 3+1 dimensional systems. One possibility is to decouple the gauge field as much as possible~\cite{Ligterink2000} and then find a suitable ansatz. Another possibility is to complement Gaussian states with unitary transformations that would respect Gauss law. On the other hand, $V_1$ and $V_2$ are also valid in higher spatial dimensions. However to decouple the $SU(N>2)$ external charges is not straightforward, since the method given in the paper only applies to SU(2). Thus, one has to investigate how to extend the method to larger $N$. Additionally, these two transformations for the SU(2) case might have also applications in condensed matter physics. For example, they could potentially be applied to study problems such as Ruderman-Kittel-Kasuya-Yosida interaction between two Kondo impurities induced by the fermionic bath. In fact, in a recent work~\cite{2018arXiv180604426G}, the spinon-holon bound state in the 2D $t-J$ model is studied from the LGT point of view, i.e., the meson formation, which provides some hints to understand high-Tc superconductivity. Nevertheless the extension to high temperatures requires the generalization of the current variational approach. Moreover a finite chemical potential that tunes the total particle number can be easily considered by just adding a new quadratic term into the Hamiltonian. 

Although here we combine the rotated Hamiltonian $H_{\Theta}$ with the use of non-Gaussian states, we expect the decoupled formulation itself to be useful for a variety of other approaches. On the one hand, as our MPS results for the SU(2) case show, this formulation can be directly addressed with tensor networks. Compared to previous MPS studies of SU(2) gauge models~\cite{kuhn,kuhnEE} we do not have to truncate the gauge Hilbert spaces to finite dimension and our formulation can be readily extended to arbitrary gauge groups SU($N$). On the other hand, it could have potential applications for the design of future quantum simulators for LGT. In the quantum simulation of the SU(2) LGT, two spins 1/2 can be considered on each site to realize the spin Hamiltonian \footnote{The spin Hamiltonian can be found in Appendix C.} via a Jordan-Wigner transformation \cite{kuhn}. This formulation generalizes the one recently implemented in a quantum simulator for the Schwinger model~\cite{Exp} to non-Abelian gauge groups. Since the gauge field can be completely eliminated, this might allow for simpler experimental realizations compared to previous proposals. The rotated Hamiltonian might also be interesting for studying the large $N$ limit of gauge theories~\cite{Hooft1974}. Since the gauge degrees of freedom are absent in the rotated Hamiltonian the effort for addressing larger values of $N$ only grows moderately.

\begin{acknowledgments}

We would like to thank the authors of Ref.~\cite{VQED} for kindly sharing with us their results in order to test the Gaussian ansatz in comparison to MPS results for the U(1) lattice gauge theory. We are also very grateful to Erez Zohar for enlightening discussions and comments.
 Moreover, PS acknowledges financial support from ``la Caixa'' fellowship grant for
post-graduate studies. TS acknowledges the Thousand-Youth-Talent Program of China. SK acknowledges financial support from Perimeter Institute for Theoretical Physics. This research was supported in part by Perimeter
Institute for Theoretical Physics. Research at Perimeter Institute is supported by the Government of Canada through the Department of Innovation, Science and Economic Development Canada and by the Province of Ontario through the Ministry of Research, Innovation and Science. ED acknowledges support from Harvard-MIT CUA, NSF Grant No. DMR-1308435, AFOSR Quantum Simulation MURI, AFOSR-MURI: Photonic Quantum Matter, award FA95501610323. JIC acknowledges ERC Advanced Grant QENOCOBA under the EU Horizon 2020 program (grant agreement 742102).
\end{acknowledgments}

\appendix
\section{Eliminating the gauge field \label{app:Elim}}
Here we explicitly derive the unitary transformation which allows for eliminating the gauge field in Hamiltonian \eqref{HU} for any non-Abelian gauge group $G$ with unitary representations as it is for example the case of SU($N$) with $N>1$. Different to derivation in Ref.~\cite{NAtrans,Lenz1994}, we choose to work with open boundary conditions. Moreover, for the ease of notation, we use the Einstein summation convention for group indices throughout this Appendix.

\subsection{Transformation to decouple the gauge field}
In order to eliminate the gauge field appearing in the hopping term of the Hamiltonian \eqref{HU}  we look for a unitary transformation given by a fermionic operator that leaves the matrix components $(U^j)_{\alpha\beta}$ invariant. Let us recall how $\phi_n^{\alpha}$ and $\phi^{\alpha,\dagger}_{n}$ transform under gauge transformations~\cite{Zohar2015b}
\begin{align}
\begin{aligned}
\Theta^{Q,j}_{g}\phi_n^{\alpha}\Theta^{Q,j\dagger}_{g}=\sum_{\beta}{D^{j}_{\alpha\beta}(g^{-1})\phi_n^{\beta}},\\ \Theta^{Q,j}_{g}\phi^{\alpha,\dagger}_{n}\Theta^{Q,j\dagger}_{g}=\sum_{\beta}{\phi_{n}^{\beta,\dagger}D^j_{\beta\alpha}(g)} ,
\end{aligned}
\label{fermtrans}
\end{align}
where $\Theta _{g}^{Q,j}=\prod_{n}{\exp (i\bm{\alpha}_{n}\cdot \bm{\mathcal{Q}}_n)}$ with $g\in G$ and $D^j(g)$ is the representation given by
\begin{equation}
D^j_{\beta\alpha}(g)=\Big( e^{i\bm{\alpha}_n\cdot\bm{T}^j}\Big)_{\beta\alpha}.
\end{equation}

From now on the representation index $j$ will be omitted as long as there is no need for a particular one. Given these transformation laws, the hopping term changes as follows
\begin{align}
&H_\mathrm{hop} \mapsto \Theta^Q H_\mathrm{hop}\Theta^{Q\dagger}\\ \notag
&=\varepsilon\sum_n{\{(\phi_n^{\dagger})_{\gamma}D_{\gamma\alpha}(g_n)(U_n)_{\alpha\beta}} D_{\beta\delta}(g_{n+1}^{-1}) (\phi_{n+1})_{\delta}+\textrm{H.c.}\},
\end{align}
where we denote the gauge transformation applied to the fermionic field living on the site $n$ by $D(g_n)$. Thus, in order to remove the gauge field in the hopping term with this transformation we see that the condition $D(g_n)U_nD(g_{n+1}^{-1})=\mathds{1}$ has to be fulfilled, or equivalently
\begin{equation}
D(g_{n+1})=D(g_n)U_n \hspace{20pt} \forall \, n.
\label{Cond}
\end{equation}

To solve the set of equations \eqref{Cond}, we fix the left open boundary conditions to $U_0=\mathds{1}$ and $D(g_0)=\mathds{1}$ for both gauge and fermionic fields, giving as a result $D(g_1)=\mathds{1}$ \footnote{Note that for periodic boundary conditions the $0$th link corresponds to the $N$th link. Hence, in that case we can use this condition to fix the boundary values.}. Solving for all links $n$ we obtain
\begin{equation}
D(g_n)=U_1U_2\cdots U_{n-1} \hspace{20pt}\forall \, n
\label{tranf}
\end{equation}
where we did not make reference to any specific representation and the index $j$ has been omitted. In fact we see that the unitary transformation $\Theta$ which generates the linear transformation \eqref{tranf} is given by
\begin{equation}
\Theta=\prod^{\rightarrow}_{k=1}{\exp\Big(i\bm{\theta}_k\cdot\sum_{m> k}{\bm{\mathcal{Q}}_m \Big)}}\equiv \prod^{\rightarrow}_{k=1}{W_k},
\end{equation}
where the link variables $\theta^a_n$ describing the gauge field were discussed below Eq.~\eqref{CR} and the superscript ``$\longrightarrow$'' means that the operators $W_k=\exp\Big(i\bm{\theta}_k\cdot\sum_{m> k}{\bm{\mathcal{Q}}_m \Big)}$ must be ordered from left to right with increasing index $k$.

\subsection{Rotated Hamiltonian}
So far we have only considered the hopping part of the Hamiltonian. To complete the transformation of the Hamiltonian, we still have to compute the mass term and the color-electric energy term. Using Eq. \eqref{fermtrans}, it is easy to see that the mass term is invariant under $\Theta$. To see how the electric field transforms under $\Theta$, we use the commutation relation between the conjugate variables from Eq. \eqref{eq:commutation_LU_RU} and then since $W^j_k=\exp\Big(i\bm{\theta}_k\cdot\sum_{m> k}{\bm{\mathcal{Q}^j}_m \Big)}$, we see that $W^j_k$ has the same matrix structure as $(U_k^j)_{mn}$ because the fermionic charges $\mathcal{Q}_m^{a,j}$ are just complex numbers on the gauge Hilbert space. Therefore we obtain
\begin{align}
W_k^jL_k^{a}W_k^{j\dagger}=L_k^{a}-\sum_{m>k}{\mathcal{Q}_m^{a,j}}
\end{align}
Thus, omitting the $j$ index, $L_n^{a}$ transforms under $\Theta$ as
\begin{align}\label{Ltransf}
\Theta L^{a}_n\Theta^{\dagger}&= L_n^{a}-\sum_{m>n}{\Big(U^{\text{Adj.}}_{n-1}\dots U^{\text{Adj.}}_1\bm{\mathcal{Q}}_m\Big)^{a}},
\end{align}
where we have applied the transformation law
\begin{equation}  W_k\mathcal{Q}_m^{a}W_k^{\dagger}=(U^{\text{Adj.}}_k)_{a,b}\mathcal{Q}^{b}_m
\end{equation}
for $m>k$ and $0$ otherwise, i.e., $\bm{\mathcal{Q}}_m$ transforms as a 3-vector under color rotations. Thus in the rotated frame Gauss' law takes the easy form
\begin{equation}
L_n^{a}=R_{n-1}^{a}\hspace{15pt} \forall \,n>1.
\end{equation}

Now using the relation between $\bm{L}=(L_{a})_{a}^{\text{dim}(G)}$ and $\bm{R}=(R_{a})_{a}^{\text{dim}(G)}$ given by
\begin{equation}
R_n^{a}=(U_n^{\text{Adj.}})_{a b}L_n^{b},
\end{equation}
with $U_n^{\text{Adj.}}$ the adjoint representation of a group element on the link, we obtain the easy form for the electric field
\begin{equation} \label{L1}
L_n^{a}=\Big[ U^{\text{Adj.}}_{n-1}\dots U^{\text{Adj.}}_1\Big(\bm{R}_0+\sum_{m\geq 1}{\bm{\mathcal{Q}}_m}\Big)\Big ]^{a}
\end{equation}
expressing $\bm{L}_n$ as a sum of the previous charges. Thus, using the orthogonality of the matrices $U_n^{\text{Adj.}}$ and combining Eqs.~\eqref{Ltransf} and \eqref{L1}, the rotated electric term takes the form
\begin{equation}
H_e=\Theta \bar{H}_e \Theta^{\dagger}=\sum_{n}{\Big(\bm{R_0}+\sum_{m\leq n}{\bm{\mathcal{Q}}_m} \Big)^2}.
\end{equation}

If the background field vanishes $\bm{R}_0=0$ and the total charge is zero, $\sum_n{\mathcal{Q}_n^{a}}=0$, which implies Gauss' law in the form of Eq.~\eqref{GU}, we can rewrite the electric term in the symmetric form

\begin{align}
\begin{aligned}
H_e=&{ \sum_{k,p}{\frac{1}{2}\sum_n\big({\theta_{kn}^{-}\theta_{pn}^{-}+\theta_{nk}^{+}\theta_{np}^{+}}\big)\bm{\mathcal{Q}}_k\bm{\mathcal{Q}}_p}}\\
=&{ \sum_{k,p}{\frac{1}{2}\sum_n\big({\theta_{kn}^{-}\theta_{pn}^{-}+\theta_{nk}^{+}\theta_{np}^{+}}-1\big)\bm{\mathcal{Q}}_k\bm{\mathcal{Q}}_p}},
\end{aligned}
\end{align}
where $\theta^{\pm}_{kn}$ are Heaviside functions defined as $\theta^{\pm}_{kn}=\theta(k-n\pm 0^+)$. Realizing that
\begin{equation}
\frac{1}{2}\sum_n\big({\theta_{kn}^{-}\theta_{pn}^{-}+\theta_{nk}^{+}\theta_{np}^{+}}-1\big)=-\frac{1}{2}\abs{k-p}
\end{equation}
is the Coulomb potential in 1+1 dimensions, we can express the Hamiltonian in the final form
\begin{align}
\begin{aligned}
H_{\Theta}=&\varepsilon\sum_n{\{ \bm{\phi}^{\dagger}_n\bm{\phi}_{n+1}+\textrm{H.c.}\}}+m\sum_n{(-1)^n\bm{\phi}_n^{\dagger}\bm{\phi}_n}\\
&+\frac{g^2a}{2}\sum_{n,m}{\bm{\mathcal{Q}}_nV_{n,m}\bm{\mathcal{Q}}_m}. 
\end{aligned}
\label{HQT}
\end{align}

Note that for the case of U(1), the previous result is much simpler, since in this case the group dimension is one and the structure constants are trivial.

\section{Effective Hamiltonian SU(2) \label{sec:SU2eff} }
As we already discussed in the main text, one needs to obtain the effective (state-dependent) single-particle Hamiltonian $\mathcal{H}(\Gamma)$ in order to solve the evolution equations \eqref{DynIm} and \eqref{DynRe} for the Gaussian state. In the following we briefly introduce the general approach and derive $\mathcal{H}(\Gamma)$ by taking derivatives of the expectation value for the rotated Hamiltonian in Eq.~\eqref{HSU2dec} with respect components of the covariance matrix $\Gamma$. In order to compute this expectation value, which involves exponential terms, we follow Appendix D in Ref.~\cite{Tao}.

\subsection{Imaginary time evolution of Gaussian states and ground state properties}
In this section, we study the zero temperature properties by assuming the ground state $\left\vert \Psi _{2}\right\rangle $ of $H_2$ as the fermionic Gaussian state $\left\vert \mathrm{GS}\right\rangle $ characterized by the covariance matrix $\Gamma _{m}=i\left\langle [A,A^{T}]\right\rangle /2$ of the Majorana fermion operator $A=(\bm{\phi} ^{\dagger }+\bm{\phi} ,i\bm{\phi} ^{\dagger }-i\bm{\phi} )^{T}$ where $\bm{\phi}^{\dagger}$ and $\bm{\phi}$ are vectors collecting all creation and annihilation operators on the lattice. The ground state properties can also be described by the covariance matrix in the Nambu basis $\Gamma=\avg{\bm{\Phi}^{\dagger}\bm{\Phi}}$ where $\bm{\Phi}=(\bm{\phi},\bm{\phi}^{\dagger})^T$. Both representations are equivalent with $\Gamma$ and $\Gamma _{m}$ related via $\Gamma=1/2-iW_{f}^{\dagger }\Gamma _{m}W_{f}/4$ with
\begin{equation}
W_{f}=
\begin{pmatrix}
1 & 1 \\
-i & i%
\end{pmatrix}
.
\end{equation}

The effective mean-field Hamiltonian $H_{Q}(\Gamma )=1/2\bm{\Phi}^{\dagger } \mathcal{H}(\Gamma )\bm{\Phi}$ can be obtained by computing the derivatives of $\left\langle H_{2}\right\rangle $ with respect to $\Gamma _{m}$~\cite{Tao}. Here, in terms of $\Gamma _{m}$ the mean values in $\left\langle H_{2}\right\rangle $ are
\begin{equation}
\left\langle P_{c}\right\rangle =(-1)^{N}\text{Pf}\left(\frac{\Gamma _{F}^{c}}{2}\right),
\end{equation}
and
\begin{align}
\begin{aligned}
\left\langle P_{c}\phi _{n\alpha }^{\dagger }\right.&\left.\phi _{n\beta }\right\rangle =\frac{1}{4}(-1)^{N}\text{Pf}\left(\frac{\Gamma _{F}^{c}}{2}\right)U_{c,\beta \delta}^{\dagger }\Bigl[\begin{pmatrix} 1 & i\end{pmatrix} (i\sigma _{y}\Gamma _{m}^{c}-1)\Bigr.\\
&\times \frac{1}{1+\frac{1}{2}(1+\sigma )(i\sigma _{y}\Gamma _{m}^{c}-1)}\sigma
\begin{pmatrix}
1 \\
-i
\end{pmatrix}
\Bigr.\Bigl]_{n\delta ,n\gamma }U_{c,\gamma \alpha },
\end{aligned}
\end{align}
where \textquotedblleft $\text{Pf\textquotedblright }$ stands for the Pfaffian, $\sigma _{y}=\tau ^{y}\otimes \mathds{1}_{2N}$, $\sigma =\mathds{1}_{2}\otimes \tau ^{z}\otimes \mathds{1}_{N}$,
\begin{align}
\begin{aligned}
\Gamma _{F}^{c}&=\sqrt{1+\sigma }\Gamma _{m}^{c}\sqrt{1+\sigma }-i\sigma
_{y}(1-\sigma )\\
\Gamma _{m}^{c}&=O_{c}\Gamma _{m}O_{c}^{T}
\end{aligned}
\end{align}
$O_c$ is a orthogonal matrix given by
\begin{equation}
O_{c}=
\begin{pmatrix}
\text{Re}\,U_{c} & -\text{Im}\,U_{c} \\
\text{Im}\,U_{c} & \phantom{-}\text{Re}\,U_{c}%
\end{pmatrix},
\end{equation}
and $U_{c}=e^{i\pi \tau ^{y}/4},$ $e^{-i\pi \tau ^{x}/4}$, $I$ for $c=x,y,z$.

Taking the derivative with respect to $\Gamma _{m}$, the single-particle Hamiltonian has the form
\begin{equation}
\mathcal{H=}
\begin{pmatrix}
\mathcal{E}_{0} & \Delta  \\
\Delta ^{\dagger } & -\mathcal{E}_{0}^{T}%
\end{pmatrix}
+i\frac{1}{2}W_{f}^{\dagger }\mathcal{H}_{P}W_{f},
\end{equation}
\begin{widetext}
	where
	\begin{align}
	\begin{aligned}
	\big(\mathcal{E}_{0}\big)_{m\alpha ,n\beta } &=\varepsilon\delta _{m,n\pm 1}\delta_{\alpha \beta }+(-1)^{n}m \delta _{\alpha \beta }\delta _{nm}+\frac{g^2a}{2}\left[\frac{3}{8}N\delta _{\alpha \beta }\delta _{nm}+\frac{1}{2}(s_{1}\tau _{\alpha \beta}^{x}{V}_{nn_{1}}+s_{2}\tau _{\alpha \beta }^{z}{V}_{nn_{2}})\delta _{nm}\right. \\
	&\Bigl. +\frac{1}{2}\sum_{k,a}\tau _{\alpha \beta }^{a}{V}_{nk}\left\langle \bm{\phi}_{k}^{\dagger }\tau ^{a}\bm{\phi}_{k}\right\rangle \delta _{nm}-\frac{1}{2}{V}_{nm}\sum_{a}\tau_{\alpha \delta }^{a}\left\langle \phi_{m\gamma }^{\dagger }\phi _{n\delta }\right\rangle \tau _{\gamma \beta}^{a}\Bigr], \\
	\big(\Delta \big)_{m\alpha ,n\beta } &=\frac{g^2a}{2}\frac{1}{2}{V}_{nm}\sum_{a}\tau_{\alpha \gamma }^{a}\left\langle \phi _{m\delta }\phi _{n\gamma}\right\rangle \tau _{\beta \delta }^{a}.
	\end{aligned}
	\end{align}
	The Hamiltonian in the Majorana basis consists of three parts, $\mathcal{H}_{P}=\frac{g^2a}{2}\Big[\mathcal{H}_{P}^{(1)}+\mathcal{H}_{P}^{(2)}+\mathcal{H}_{P}^{(3)}\Big]$,  with
	\begin{align}
	\mathcal{H}_{P}^{(1)} =&{V}_{n_{1}n_{2}}\left[s_{1}s_{2}\left\langle P_{x}\right\rangle O_{x}^{T}\sqrt{1+\sigma }\frac{1}{\Gamma _{F}^{x}}\sqrt{1+\sigma }O_{x}
	-i^{\mathcal{N}}s_{2}\left\langle P_{y}\right\rangle O_{y}^{T}\sqrt{1+\sigma }\frac{1}{\Gamma _{F}^{y}}\sqrt{1+\sigma }O_{y}-s_{1}\left\langle P_{z}\right\rangle \sqrt{1+\sigma }\frac{1}{\Gamma _{F}^{z}}\sqrt{1+\sigma }\right],
	\end{align}
	\begin{align}
	\begin{aligned}
	\big(\mathcal{H}_{P}^{(2)}\big)_{ij} =\sum_{n}{V}_{nn_{1}}&\left\{-(s_{1}s_{2} \tau ^{z}-is_{2}\tau ^{y})_{\alpha \beta }\left\langle P_{z}\phi _{n\alpha}^{\dagger }\phi _{n\beta }\right\rangle \left(\sqrt{1+\sigma }\frac{1}{\Gamma _{F}^{z}}\sqrt{1+\sigma }\right)_{ij} -i\frac{1}{2}(-1)^{N}\text{Pf}\left(\frac{\Gamma _{F}^{z}}{2}\right)\right.\\
	&\left.\phantom{-}\times\left[\frac{1}{1+\frac{1}{2}(1+\sigma )(i\sigma _{y}\Gamma _{m}-1)}
	\begin{pmatrix}
	1 \\
	-i%
	\end{pmatrix}
	\right]_{j,n\alpha }  (s_{1}s_{2}-s_{2}\tau ^{x})_{\alpha \beta }\left[\begin{pmatrix} 1 & i\end{pmatrix} \frac{1}{1+%
		\frac{1}{2}(i\Gamma _{m}\sigma _{y}-1)(1+\sigma )}\right]_{n\beta ,i}\right\},
	\end{aligned}
	\end{align}%
	and
	\begin{align}
	\begin{aligned}
	\big(\mathcal{H}_{P}^{(3)}\big)_{ij} =\sum_{n}{V}_{nn_{2}}&\left\{-(-s_{2}\tau^{x}-i\tau ^{y})_{\alpha \beta }\left\langle P_{x}\phi _{n\alpha }^{\dagger}\phi _{n\beta }\right\rangle \left(O_{x}^{T}\sqrt{1+\sigma }\frac{1}{\Gamma_{F}^{x}}\sqrt{1+\sigma }O_{x}\right)_{ij} -i\frac{1}{2}(-1)^{N}\text{Pf}\left(\frac{\Gamma _{F}^{x}}{2}\right) \right.\\
	&\times\left[O_{x}^{T}\frac{1}{1+\frac{1}{2}(1+\sigma )(i\sigma _{y}\Gamma _{m}^{x}-1)}
	\begin{pmatrix}
	1 \\
	-i%
	\end{pmatrix}
	\right]_{j,n\gamma }
	\gamma U_{x,\gamma \alpha }(-s_{2}\tau ^{x}-i\tau ^{y})_{\alpha \beta}U_{x,\beta \delta }^{T}\\
	&\left.\times\left[\begin{pmatrix} 1 & i\end{pmatrix} \frac{1}{1+\frac{1}{2}(i\Gamma_{m}^{x}\sigma _{y}-1)(1+\sigma )}O_{x}\right]_{n\delta ,i}\right\},
	\end{aligned}
Es	\end{align}
\end{widetext}
The imaginary time evolution of $\Gamma $ obeys Eq. \eqref{DynIm} which yields the ground state configuration in the limit $\tau \rightarrow\infty $. The variational state in different sectors reads%
\begin{align}
& \frac{1}{4\sqrt{2}}\Theta ^{\dagger }\Big[(\ket{\uparrow}_{z}+s_{1}%
\ket{\downarrow}_{z})\big[(1+s_{2})\ket{\uparrow}_{z}+(1-s_{2})%
\ket{\downarrow}_{z}\big] \notag  \label{CA} \\
& +s_{2}(s_{1}\ket{\uparrow}_{z}-\ket{\downarrow}_{z})\big[(1+s_{2})%
\ket{\uparrow}_{z}+(1-s_{2})\ket{\downarrow}_{z}\big]P_{z}
\notag \\
& -s_{2}(\ket{\uparrow}_{z}+s_{1}\ket{\downarrow}_{z})\big[(1-s_{2})%
\ket{\uparrow}_{z}+(1+s_{2})\ket{\downarrow}_{z}\big]P_{x}
\notag \\
& +(s_{1}\ket{\uparrow}_{z}-\ket{\downarrow}_{z})\big[(1-s_{2})\ket{\uparrow}%
_{z}+(1+s_{2})\ket{\downarrow}_{z}\big]i^{\mathcal{N}}P_{y}%
\Big]\notag \\
&\times \ket{\text{GS}}\left\vert 0\right\rangle _{\text{gauge}}
\end{align}

In fact, considering as Gaussian state $\ket{\text{GS}}$ the one corresponding to the Dirac sea in Eq.~\eqref{eq:SC_SU2}, the previous ansatz acquires the simple form
\begin{align}
&\Theta ^{\dagger }\frac{1}{\sqrt{2}} |\psi_{\text{SC},\text{SU(2)}}\rangle
\frac{1}{2}[(1+s_{1})\left\vert \uparrow \right\rangle _{z}\left\vert
\uparrow \right\rangle _{z}+s_{2}(s_{1}-1)\left\vert \uparrow \right\rangle
_{z}\left\vert \downarrow \right\rangle _{z}  \notag \\
&+(s_{1}-1)\left\vert \downarrow \right\rangle _{z}\left\vert \uparrow
\right\rangle _{z}-s_{2}(s_{1}+1)\left\vert \downarrow \right\rangle
_{z}\left\vert \downarrow \right\rangle _{z}]\left\vert 0\right\rangle _{\text{gauge}}
\end{align}%
since the Dirac sea is a common eigenstate of the parity operators $P_{c}$. For instance, in the subspace with $s_{1}=s_{2}=-1$, the initial seed state
\begin{equation}
\Theta ^{\dagger }\frac{1}{\sqrt{2}} |\psi_{\text{SC},\text{SU(2)}}\rangle [%
\ket{\uparrow}_{z}\ket{\downarrow}_{z}-\ket{\downarrow}_{z}\ket{\uparrow}%
_{z}]\left\vert 0\right\rangle _{\text{gauge}}
\end{equation}%
describes the singlet state of static charges in the (deep) string regime.

\section{Technical details of the MPS simulations\label{app:mps}}
In order to solve the SU(2) LGT with MPS we start from the decoupled Hamiltonian $H_{\Theta }$ from Eq. \eqref{H}. For convenience in the simulations we chose to translate the fermionic degrees of freedom to  spins via a Jordan Wigner transformation~\cite{kuhn}. The resulting Hamiltonian reads
\begin{align}
\begin{aligned}
H&=\varepsilon\sum_{n}{\big(\sigma _{r,n}^{+}\sigma _{g,n}^{z}\sigma
	_{r,n+1}^{-}+\sigma _{g,n}^{+}\sigma _{r,n+1}^{z}\sigma _{g,n+1}^{-}+\mathrm{%
		H.c.}\big)}  \\
& +m\sum_{n}{(-1)^{n}\big(2+\sigma _{n,r}^{z}+\sigma _{n,g}^{z}\big)}+\frac{g^2a}{2}H_{e}
\end{aligned}
\label{spinH}
\end{align}
where the $\sigma$-matrices are the usual Pauli matrices, the electric term $H_e$ takes the form given in Eq.~\eqref{He}, and the subscript indicates the vertex and the color on which they are acting. The dynamic charges in spin formulation are given by~\cite{kuhn}
\begin{align}
Q_{n}^{x}&=-\frac{i}{2}\big(\sigma _{r,n}^{+}\sigma _{g,n}^{-}-\text{H.c.}\big), \\
Q_{n}^{y}&=-\frac{1}{2}\big(\sigma _{r,n}^{+}\sigma _{g,n}^{-}+\text{H.c.}\big), \\
Q_{n}^{z}&=\frac{1}{4}\big(\sigma _{r,n}^{z}-\sigma _{g,n}^{z}\big).
\end{align}
In our calculations we are interested in the subsector of vanishing total charge which we ensure by adding the energy penalty $\lambda ( \sum_{n}{\bm{\mathcal{Q}}_n})^2$ to the Hamiltonian. The constant $\lambda$ has to be chosen large enough to sufficiently penalize states with nonzero charge. The ground state of the resulting Hamiltonian can be computed with standard variational optimization of the MPS wave function~\cite{Verstraete2008,Orus2014}. To estimate our numerical errors, we run our simulation for a series of bond dimensions $\chi$ of the MPS ranging from $40$ up to $200$ and values of $\lambda$ up to $5000$. The MPS results presented in the main text correspond to $\chi=200$, $\lambda=1000$ for which we find that both our numerical errors in the ground state energy and the correlation function as well as the expectation value of the penalty are negligible.

\bibliography{References_prd}

\end{document}